\documentclass[preprint]{aastex631}

\newcommand{\eg}{{\it e.g.}}

\newcommand{\al}{\mbox{ \raisebox{-.8ex}{$\stackrel{\textstyle <}{\sim}$} }}

\usepackage{natbib}
\usepackage{color}
\usepackage{graphicx}
\usepackage{grffile}
\usepackage{amsmath, amsthm, amssymb}
\usepackage{amssymb}
\usepackage{ulem}
\usepackage{soul}

\begin{document}
\shorttitle{Transport in Gravitationally Unstable Disks}
\shortauthors{Steiman-Cameron et al.}

\title{Mass and Angular Momentum Transport in a Gravitationally Unstable Protoplanetary Disk with Improved 3D Radiative Hydrodynamics}

\correspondingauthor{Thomas Y. Steiman-Cameron}
\email{tomsc@astro.indiana.edu}.

\author{Thomas Y. Steiman-Cameron}
\affiliation{Astronomy Department, Indiana University, Bloomington, IN 47405, USA}

\author{Richard H. Durisen }
\affiliation{Astronomy Department, Indiana University, Bloomington, IN 47405, USA}

\author{Aaron C. Boley}
\affiliation{Dept. of Physics and Astronomy, University of British Columbia, Vancouver, BC V6T 1Z1, Canada}

\author{Scott Michael}
\affiliation{Astronomy Department, Indiana University, Bloomington, IN 47405, USA}

\author{Karna Desai}
\affiliation{Astronomy Department, Indiana University, Bloomington, IN 47405, USA}

\author{Caitlin R. McConnell}
\affiliation{Astronomy Department, Indiana University, Bloomington, IN 47405, USA}

\begin{abstract}

During early phases of a protoplanetary disk’s life, gravitational instabilities (GIs) can produce significant mass transport, can dramatically alter disk structure, can mix and shock-process gas and solids, and may be instrumental in planet formation. We present a 3D grid-based radiative hydrodynamics study with varied resolutions of a 0.07 M$_\odot$ disk orbiting a 0.5 M$_\odot$ star as it settles over most of its radial extent into a quasi-steady asymptotic state that maintains approximate balance between heating produced by GIs and radiative cooling governed by realistic dust opacities. We assess disk stability criteria, thermodynamic properties, strengths of GIs, characteristics of density waves and torques produced by GIs, radial mass transport arising from these torques, and the level to which transport can be represented as local or nonlocal processes. Physical and thermal processes display distinct differences between inner optically thick and outer optically thin regions of the disk. In the inner region, gravitational torques are dominated by low-order Fourier components of the azimuthal mass distribution. These torques are strongly variable on the local dynamical time and are subject to rapid flaring presumably driven by recurrent swing amplification. In the outer region, $m=1$ torques dominate. Ring-like structures exhibiting strong noncircular motions, and vortices develop near the inner edge between 8 and 14 au. We find that GI-induced spiral modes erupt in a chaotic manner over the whole low-$Q$ part of the disk, with many spiral modes appearing and disappearing, producing gravitoturbulence, but dominated by fluctuating large-scale modes, very different from a simple $\alpha$-disk. 

\end{abstract}

\section{Introduction}

Recent capabilities provided by the Atacama Large Millimeter/submillimeter Array (ALMA) and other instruments have revealed beautifully detailed structures, including rings, spiral arms, and forming gas giants, in protoplanetary disks (PPDs)
\citep{ALMA_Partnership2015, Huang_etal_2017ApJ, Huang_etal_2018_ApJ_DSHARP.II, Huang_etal_2018_ApJ_DSHARP.III, Huang_etal_2020ApJ_GMAur, Paneque-Carreno2021ApJ, Currie_etal_2022}. 
During the early phases (Class 0 and 1) of a PPD's life, gravitational instabilities (GIs) can produce significant mass transport, 
dramatically alter disk structure, mix and shock-process gas and solids, and be instrumental in planet formation. 

The existence, persistence, and characteristics of GIs in PPDs are driven  primarily by thermal and accretion processes.  When cooling is sufficiently rapid, fragmentation can occur \citep{Boss1997, gammie2001, mayer2002}.
Such disk fragmentation can further lead to gravitationally bound clumps, which themselves offer a wide range of evolutionary possibilities. For example, the clumps can undergo nontrivial disk migration \citep{Vorobyov_Basu_2010,Baruteau_etal_2011, Michaelphd2011, Michael_Durisen_Boley2011}, grow to become brown dwarfs or stars \citep{Kratter_etal_2010ApJ708,Kratter_etal_2010ApJ710}, or remain as giant planets \citep{boley2010}. Moreover, clumps that are initially bound are not guaranteed to remain so, and through disk migration, scattering, or clump-clump interactions, they may become tidally disrupted \citep{boley2010, Nayakshin_2010}. This disruption could in turn serve as a mechanism for processing solids.
For lower cooling rates, GI-active disks can instead develop quasi-steady structures or bursts of activity that dramatically transform the disk without forming fragments.  
 For reviews, see \citet{Durisen_2011}, \citet{Armitage2011ARAA}, \citet{2016-ARAA-Kratter-Lodato}, \citet{Rice2016}, and \citet{Armitage2019}.
 
Detailed three-dimensional (3D) hydrodynamic modeling of GI-active PPDs, as reported herein, can provide considerable insight into the physical and thermal states of these disks and their short- and long-term evolution.  An important consequence of GIs in nonfragmenting PPDs is the ability of GI-generated spiral arms to drive angular momentum transport, a fact initially recognized by \citet{lyndenbell1972} 
in the context of galactic dynamics.  A disk's susceptibility to GIs can be parameterized by the
Toomre  $Q$ parameter, 
\begin{equation}
Q=c_s\kappa / \pi G \Sigma,
\end{equation}
where $c_s$ is the sound speed, $\kappa$ is the epicyclic frequency, and $\Sigma$ is the disk surface 
mass density \citep{toomre1981}.   In a disk subject to GIs, 
small density perturbations grow exponentially on a timescale comparable to the local dynamical time
when $Q \al$ 1.7 in the {\it linear regime} \citep{durisen2007}. In the {\it nonlinear regime}, perturbations can grow for even larger $Q$.   These
perturbations manifest themselves  as predominantly trailing  multiarm spiral density waves
over a broad range of radii; see \cite{papaloizou1991}, \citet{laughlin1998}, \citet{nelson1998},
\citet{nelson2000}, \citet{pickett1998}, \citet{pickett2000}, \citet{pickett2003}, and \citet{Michael2012}.  

Gravitational torques arising from these spiral structures 
enable the disk to tap the free energy associated with the rotational shear.   
Some of this energy is then returned as heat when waves steepen into shocks.
This heating, along with net inward transport of mass, 
pushes the disk back toward stability.  At the same time, radiative cooling acts in the opposite sense, pushing the disk toward continued instability.

Early studies suggested 
that the amplitude of GIs in accretion disks can self-limit, reaching a balance between radiative cooling and disk heating
at a relatively constant, but unstable, value of $Q$  \citep[e.g.,][]{goldreich1965, paczynski1978, pringle1981,  lin1987}.   Numerical simulations of PPDs have verified the existence of these limiting ``saturated GIs'' where heating and cooling balance to produce asymptotic states
for various ranges of parameter space \citep[e.g.,][]{tomley1991,tomley1994,pickett1998,pickett2000, pickett2003,gammie2001,boss2002,rice2003b,lodato2004,rice2005,mejia2005,boley2006,boleyphd2007, stamatellos2008,cossins2009, Michael2012, zhu_etal2012ApJ,Bethune_Latter_Kley2021}. 

Spiral structure is not the only type of morphology expected in disks. Indeed, while it is well established that rings can arise from disk interactions with embedded objects \citep{goldreich1980,lin1986,paardekooper2006,Zhuetal_2011ApJ,Zhangetal_2018ApJ}, 
they can also emerge in a disk without an embedded perturber 
\citep[\eg,][]{Lubow_1991ApJ, Ogilvie_2001MNRAS, Takahashi_2014ApJ, Tominaga_etal2019, Lee_etal_2019a, Lee_etal_2019b, Riols_Lesur_2019AA, Riolsetal_2020AA,Ring_Formation_LiEtal_2021ApJ}. 
In particular, eccentric modes, corresponding to perturbations with azimuthal wavenumber $m = 1$, have received particular interest in the context of PPDs because of their global nature. 
A large corpus of work has examined the development and sustenance of these modes in fluid disks \citep{AdamsRuden1989,Shu_etal_1990ApJ,Hirose_Osaki1990, Lubow_1991ApJ, Heemskerk_etal1992, Laughlin_Korchagin_1996, Ogilvie_2001MNRAS,Tremaine2001AJ, Papaloizou_2002AA, Tominaga_etal2019, Lee_etal_2019a, Lee_etal_2019b,Ring_Formation_LiEtal_2021ApJ, Bethune_Latter_Kley2021}.  It has been shown that almost any disk with a realistic density profile can sustain long-lived eccentric modes \citep{Lee_etal_2019b}.  Moreover, these modes can grow in amplitude via the sling mechanism that amplifies an eccentric perturbation through the wobble of the central star 
\citep{AdamsRuden1989, Shu_etal_1990ApJ,Lin_2015MNRAS.448.3806L}.  Ring formation can follow via angular momentum exchange with the unstable eccentric mode 
\citep{Lubow_1991ApJ,Ogilvie_2001MNRAS,Lee_etal_2019a,Lee_etal_2019b}.  A recent 2D study of an eccentric spiral instability in a self-gravitating disk with prescribed cooling by \citet{Ring_Formation_LiEtal_2021ApJ}  found that a trapped one-arm instability forms early in the simulation and evolves into a set of axisymmetric rings.   

For these reasons, we expect that rings formed by a process not involving embedded objects may be a common product of early PPD evolution. \citet{Durisen_etal_2005Icarus}, in an analysis of ring-like structures that developed in their PPD hydro simulation, proposed a hybrid avenue for planet formation where even if instabilities due to disk self-gravity do not produce gaseous protoplanets directly, they may create persistent dense rings that are conducive to accelerated growth of gas giants through core accretion. They suggested that even if GIs do not lead to permanent clump formation, they may significantly accelerate core accretion by creating persistent dense gas rings near boundaries between GI-active and GI-inactive regions \citep[see also][]{hag2003a,hag2003b}.

Ring features have been seen in previous 3D hydro simulations of self-gravitating PPDs carried out by our group and collaborators
\citep{pickett1996, pickett2003,  mejiaphd2004,  mejia2005, Durisen_etal_2005Icarus, caiphd2006,  boley2006, boley2007_bdnl, cai2006, cai2008, Michael2012, Steiman2013, desai2019}.  These studies found that rings form early in the simulation, well before disks settle into an asymptotic state where heating and cooling balance.   The simulations described in the current work also lead to multiple rings, which we explore in \S\ref{subsection:rings}.

Another important question arises if a GI-active disk settles into a quasi-steady saturated state.  Specifically, in this state does thermal balance of heating and cooling apply locally or only in a global, long-term, time-average sense?  Torques due to spiral waves involve long-range interaction for low-order (few-armed) 
spirals, and the wave nature of GI activity opens the possibility for wave 
transport of energy \citep{laughlin1996,balbus1999}. On the other hand, it has been proposed
by several authors that GI transport and evolution can be captured by a turbulent 
$\alpha$-disk formulation by applying a \citet{shakura1973} $\alpha$ that
can be obtained in the case that the energy balance is precisely local; see \citet{paczynski1978}, \citet{pringle1981}, \citet{gammie2001}, \citet{vorobyov2010}, and \citet{Zhu_etal_2010ApJ}. 
In his razor-thin shearing box simulations,
\citet{gammie2001} found good agreement between such a local derivation and the
effective $\alpha_{\rm eff}$ computed from the observed stresses in his 
simulations. However, these calculations were local by their very nature. Full 3D hydrodynamics 
calculations have given somewhat mixed results on this issue depending on the
disk mass, numerical resolution, numerical techniques, 
and the nature of the cooling \citep[e.g.,][]{lodato2004, lodato2005,boley2006, cossins2009, forgan_etal2011,michael2010,Michael2012,Steiman2013,Evans_etal_2015,Bethune_Latter_Kley2021}. 

To address further some of the issues outlined above, we report  here results of a grid-based, finite-difference, 3D radiative hydrodynamics convergence study of a gravitationally unstable PPD where cooling is allowed to adjust naturally by radiative transport using realistic dust opacities and where star-disk interactions are explicitly modeled.  This study builds on the earlier works of \citet{Michael2012} (hereafter Paper I) and \citet{Steiman2013} (hereafter Paper II) using code improvements in radiative transfer and the incorporation of star-disk interactions.  
The number of azimuthal elements used in calculations is especially important because it is nonaxisymmetric structures, i.e., spiral waves, that produce the gravitational torques
leading to mass and angular momentum transport. 
We follow the time evolution of four simulations of a PPD which differ only in the azimuthal resolution of their computational grid, allowing us to ensure that results have converged to a solution that is not affected by the size of the azimuthal mesh. 

 Simulations are run for a time period sufficient for the disk to settle into a long-lived, statistically quasi-steady, asymptotic state, allowing GIs and their consequences to be characterized and quantized. ``Quasi-steady, asymptotic state'' refers here to the evolutionary phase of a gravitationally unstable disk that has settled into a long-lived, quasi-steady balance between radiative cooling and disk heating at a relatively constant, but unstable, value of $Q$.

Paper II and this work examined the same disk with the same initial conditions and hydrodynamics code, but here with important improvements to the code (Section 2), including the implementation of a subcyling approach to better control heating and cooling and the inclusion of an indirect potential approach to self-consistently account for star -- disk interactions.  In contrast to Paper II, great convergence is found here, demonstrating the importance of doing the radiative physics well.

The balance of this paper is organized as follows.  
Section 2 provides the details of the numerical approach and defines the models.  
Results of the simulations are reported in Section 3, with physical and thermodynamic properties of the converged asymptotic disk described in Sections 3.1, 3.2, and 3.3; the time dependence of 
nonaxisymmetric modal properties presented in Section 3.4; computation of the gravitational torques arising from these structures are examined in Section 3.4; and angular momentum transport, mass flux, and time variability are described in Sections 3.5 and 3.6.  The locality/nonlocality of mass transport due to GIs and the applicability of an effective $\alpha$-based viscosity is discussed in Section 3.7.  The development of ring-like structures and their impact on disk evolution is found in Section 3.8.  Convergence is discussed in Section 3.9, followed by a discussion section in Section 4. A summary and conclusions are found in Section 5.

\section{Approach}

\subsection{Hydrodynamics} \label{subsection:hydrodynamics}
 
We seek to understand the physical and dynamical characteristics of a resolution-independent PPD simulation at a time when the disk has relaxed into a quasi-steady asymptotic state characterized by a statistically constant unstable $Q(r)$.   To this end, we conduct multiple hydrodynamic simulations of a PPD using CHYMERA,  an explicit, second-order, 3D Eulerian code that self-consistently solves the hydrodynamic equations of motion, Poisson's equation, and the energy equation, on a uniform cylindrical grid \citep{boleyphd2007, boley2007_bdnl}.  The number of grid elements in the $r$-, $z$-, and 
$\phi$-directions are specified by $j_{\rm max}$, $k_{\rm max}$, and $l_{\rm max}$, respectively, and mirror symmetry is assumed about the equatorial plane. 

CHYMERA uses a Norman-Winkler artificial viscosity to mediate shocks. Source and flux terms are computed separately in an explicit, second-order time integration \citep{norman1986,albada1982,Williams_1988CeMec,christodoulou1991,yangphd1992}.  The code has been used in several previous studies of protostellar disks conducted by the Indiana University Hydrodynamics Group and collaborators \citep[e.g.,][]{boleyphd2007,boley2007_bdnl, boley2008,boley2009,boley2010,cai2010,Michael2012,Steiman2013,Evans_etal_2015,desai2019}.

The equation of state used in this work takes into account contributions of the translational, rotational, and vibrational states of H$_2$ and assumes a fixed H$_2$ ortho-to-para ratio of 3:1 \citep{boley2007_h2}. 
For the temperature range in our simulations (Section \ref{subsection:heat_cool}), the gas is well approximated by an adiabatic index $\gamma = 5/3$.

\subsection{Radiative Cooling} \label{subsection:cooling}

The disk is embedded in a 3 K background, a boundary condition in the radiative scheme for the z-direction of the hydro code for the simulations run here.
 Disk cooling is implemented using the radiative energy transfer scheme of \cite{boley2007_bdnl} that combines flux-limited diffusion for optically thick regions in the $r$- and $\phi$- directions and a single-ray discrete-ordinate radiative transfer solver in the $z$-direction that treats both optically thick and thin regions.  This scheme produces smooth temperature profiles across the disk photosphere.   The opacity tables of 
\citet{dalessio2001} are used, assuming minimum and maximum grain sizes of 0.005 and 1.0 $\mu$m, respectively, and a power-law size distribution with index -3.5  within this range \citep[for opacity details, see Appendix A of][]{boley2006}.    

Because the code explicitly solves for radiative transport, it can become unstable if the radiative time step becomes smaller than the hydrodynamic time step producing excessive heating or cooling of the gas that lead to unphysical results and numerical instability.  To avoid this situation, in Paper II limiters were placed on the
local cooling and heating rates to prevent the computational time step $\Delta t$  from becoming shorter than $\sim 0.03$ outer rotation periods (ORPs), where ORP is defined as the initial orbital period at radial grid element $j = 200$ ($r \approx 33$ au, 1 ORP $\approx$ 255 yr; see Section \ref{subsection:indirect potential}).  While this eliminated the numerical instability, it artificially set a computational time step that might be unrealistically large.  In the current work, a subcycling approach is used to control heating and cooling.  At each hydrodynamic time step, the radiative routine calculates a separate radiative time step size from the ratio of the internal energy density to the divergence of the radiative flux for each cell and compares this with the hydrodynamic time step.  If the radiative time step size is smaller than the hydrodynamic time step, radiative transfer is subcycled, i.e., multiple calls are made to the radiative transfer routine for that hydrodynamic time step.  Details are described in \citet{shabram2013} and \citet{Evans_etal_2015}. 

\cite{boley2007_bdnl} demonstrated that, for numerical stability and
accuracy, the optically thick portion of the disk must be resolved by a minimum of about five to seven vertical cells. When the condition is not satisfied, vertical oscillations that are purely numerical can occur that could, in turn, lead to ``artificial heating'' by an uncontrolled numerical effect.  The vertical resolution of our simulations is too small to satisfy this requirement interior to $\sim$ 8 au. Thus, we lack confidence in the simulations in this region. 

\subsection{Star--Disk Interactions} \label{subsection:indirect potential}

In Papers I and II, the star was represented by a point-source gravitational field held fixed at the center of the computational grid and star-disk interactions were not modeled.  In fact, star-disk interactions will inevitably displace the star from the system center of mass \citep{rice2003b}.  Here we account for the star's acceleration using the indirect potential method \citep{AdamsRuden1989,NelsonPapaloizou00}, as discussed in \citet{michael2010}.  In this approach, the star remains fixed at the grid center for computational convenience while the reference frame of the star plus grid is accelerated through inclusion of fictitious forces that self-consistently account for the gravitational interactions between the star and disk.

\subsection{Initial Conditions} \label{subsection:Initial Conditions}

To assess mesh convergence, four simulations were run following the evolution of a 0.07 $M_\odot$ PPD surrounding a 0.5 $M_\odot$ central star with a background temperature of 3 K. These simulations have identical initial conditions and differ only in the azimuthal resolution of the  computational grid.   The numbers of grid elements in the $r$- and $z$-directions are given by $j_{\rm max} = 512$ and $k_{\rm max} = 64$, respectively, for all simulations with each increment in $j$ and $k$ corresponding to 0.167 au.  This $j_{\rm max}$ provides a sufficiently large outer computational grid radius to keep all material on the grid when the disk expands during the violent onset of nonlinear GIs and throughout the entire disk evolution.
 The number of grid elements in the $\phi$-direction is given by $l_{\rm max} = $ 64, 128, 256, or 512.   Simulations will hereafter be referenced by their $l_{\rm max}$.  

Outflow conditions are used for the vertical and radial boundaries. These are chosen, as opposed to allowing inflow and outflow, to ensure that artificial mass streams do not form at the boundaries during fluxing. Mass that flows out of the grid is removed from the simulation.  In the vertical direction, only the top boundary uses the outflow condition, while the midplane boundary assumes mirror symmetry.

Like the outer radius, the inner radius also assumes an ``outflow'' condition. However, for the inner boundary, instead of removing mass from the grid entirely, mass that passes through the inner boundary is added to the star’s mass, thus assuming that accretion has taken place. 

Time is expressed in units of ORP, defined by the orbital period in the initial $(t= 0)$ disk at radial grid element $j = 200$ $(r \approx 33$ au); 1 ORP $\approx$  255 years.  Simulations were followed through $\sim 20$ ORPs, a time when the disk has settled into a quasi-steady thermodynamic state where cooling and heating are in balance and GI-induced structural and thermal properties are approximately constant. 

The initial disk configuration that serves as the basis for these simulations was developed by 
\cite{mejia2005}; 
\citep[see also][]{pickett1996, pickett2003,  mejiaphd2004,  mejia2005, caiphd2006}. 
At $t = 0$ the initial state of the Mej\'{\i}a disk is esentropic with inner and outer radii of 2.3 and 40 au, respectively, and a surface density profile $\Sigma \propto r^{ -1/2}$ within this radial range.   The initial thermodynamic state of this disk was set using an equilibrium star plus disk model generated by a modified \citet{hachisu1986} self-consistent field relaxation method, where random $\sim 10^{-4}$ cell-to-cell density perturbations were introduced to seed the growth of GIs.    The  Mej\'{\i }a $t = 0$ disk has defined the initial state of disks in a number of hydrodynamical studies \citep[e.g.,][]{mejia2005, boley2006, boley2007_bdnl, cai2006, cai2008, Michael2012,Steiman2013, desai2019}.  In these studies, the unstable disk passes through several phases of evolution during which the disk's mass distribution and thermal state are significantly modified.  Ultimately, the disk settles into a quasi-steady, long-lived {\it asymptotic state} of sustained GI activity over a large part of the disk, with an approximate overall balance between heating and cooling.   Details of the asymptotic phase and resultant asymptotic disk structures in these works are strongly dependent on the thermodynamical properties, i.e., heating and cooling, of the disk.  

The Paper II simulations used the state of the Mej\'{\i}a disk as its initial condition, defined at $t=0$ ORPs.  The calculations reported here begin with initial conditions defined by the Paper II disk state at 7 ORPs, at the time when the disk is still in transition toward its asymptotic phase.  
The star's motion was not accounted for up until this point, but is in the calculations that follow.  This is part of the reason that we allow a lot of time for transients to decay.  Specifically, we follow the evolution of these disks through $t = 20$ ORPs for all $l_{\rm max}$ but limit most of our analyses to $t > 16$ ORPs to allow transients to fully decay and the disk to fully settle into an asymptotic phase.  As seen in the following sections, all four simulations settle into an asymptotic phase by $\sim$ 16 ORPs.

\section{Simulations}

\subsection{The Asymptotic Converged Disk}  \label{subsection:converged_disk}

Our goal is to understand the ``asymptotic converged disk configuration'' of a gravitationally unstable PPD and how this drives disk evolution. ``Asymptotic'' refers here to the evolutionary phase of a gravitationally unstable disk that has settled into a long-lived, quasi-steady balance between radiative cooling and disk heating at a relatively constant, but unstable, value of $Q$ (sometimes referred to as ``saturated GIs").  Four simulations, each with differing azimuthal resolution, are followed to their own asymptotic states.  The detailed configurations of these four asymptotic states are then used to establish grid convergence of the asymptotic disks and define the asymptotic converged disk configuration (hereafter ACDC).


\begin{figure}[htb!]
\hspace{0.5in}
\includegraphics[width=0.9\textwidth]{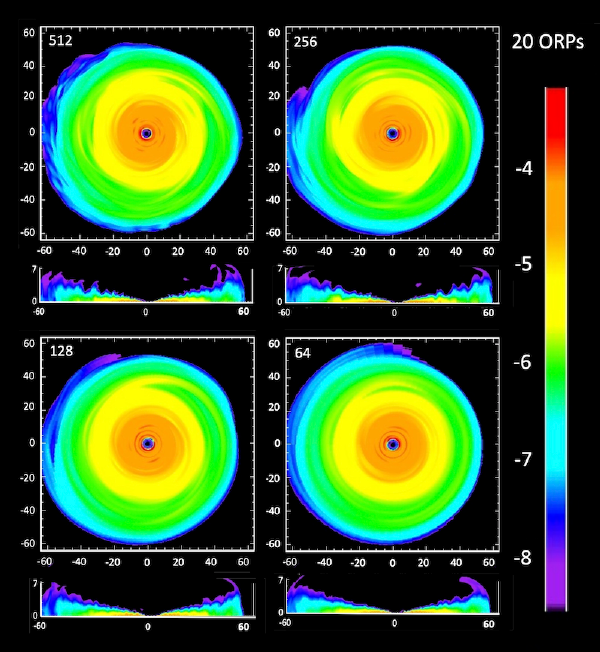}
\caption{Volume mass densities in the midplane (top of each panel) and out of the plane along an azimuthal cut (bottom of each panel) at $t = 20$ ORPs for each of the four simulations.  Panels are labeled with $l_{\rm max}$, the number of azimuthal grid elements used in the simulation.  The color scale is logarithmic in code units, and axes units are given in au. 
}
\label{fig:2DSigmaPlot}
\end{figure}


\begin{figure}[htb!]
\vspace{-10pt}
\centerline{\includegraphics[width=0.8\textwidth]{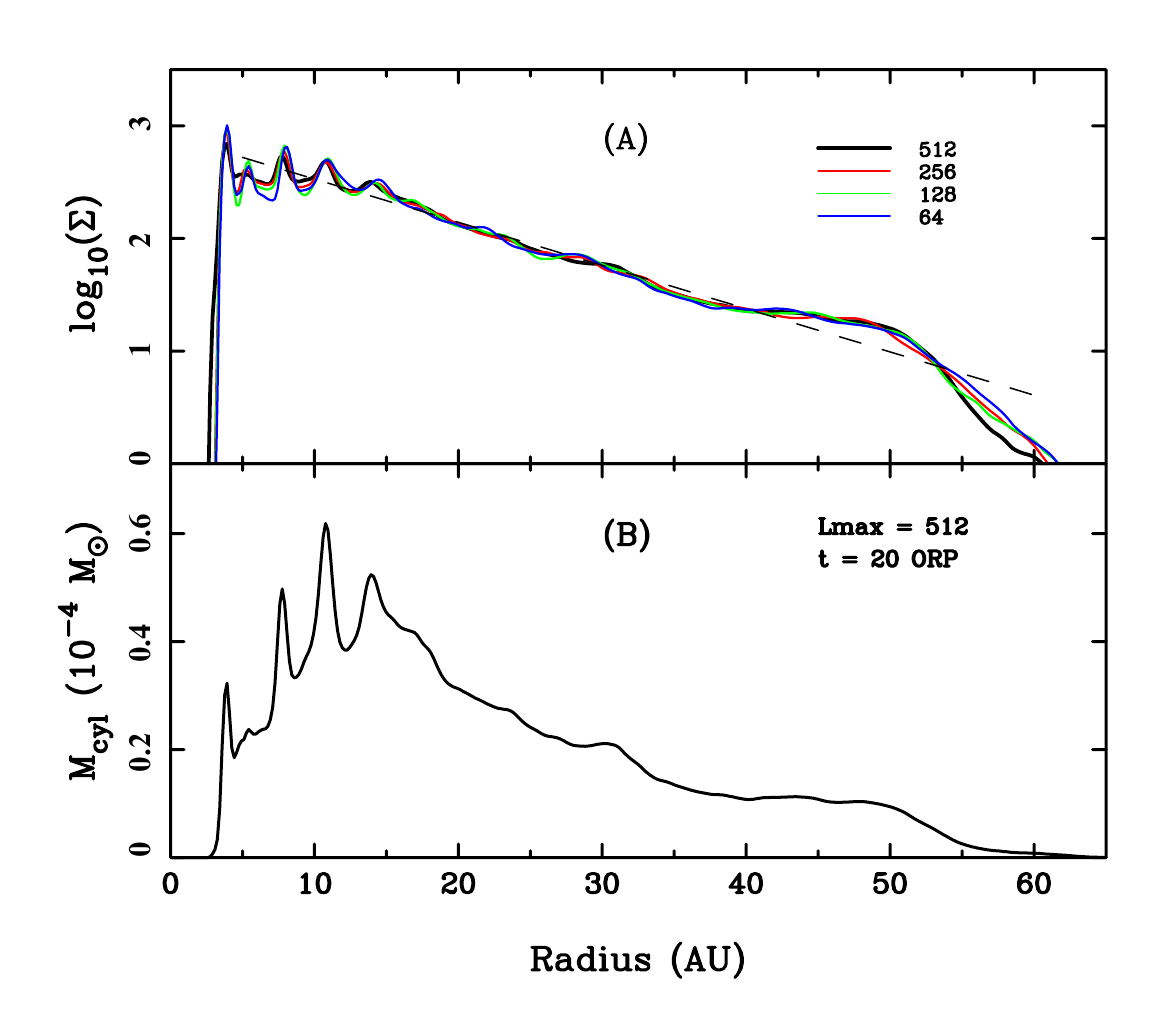}}
\caption{(a) Azimuthally averaged surface densities at $t = 20$ ORPs, measured in g ${\rm cm}^{- 2}$, as a function of radius for the $l_{max} = $ 512, 256, 128, and 64 simulations.  The dashed line depicts the best-fit exponential to the 512 $\Sigma(r)$ between 8 and 40 au.
(b) Masses on cylindrical shells one radial cell width wide (0.1667 au) at $t = 20$ ORPs for $l_{max} = 512$. Note ring-like features at 8, 11, and 14 au and broad bumps around 31--32 and 48 au. } 
\label{fig:sigmaplot}
\end{figure}

Figure~\ref{fig:2DSigmaPlot} shows mass densities in the disk midplane (top panel) and above the midplane along an azimuthal cut through the midplane (bottom panel) at $t = 20$ ORPs for all four simulations.  As described below, by this time each simulated disk has settled into a quasi-steady state.  
Figure~\ref{fig:sigmaplot}(a) shows azimuthally averaged radial surface density profiles, 
$\Sigma(r)$, for each $l_{\rm max}$ simulation at $t=20$ ORPs.  These profiles are very similar between $\sim$ 7.5 and 52 au and well described between 8 and 40 au by the exponential fit to the $l_{\rm max}$ = 512 profile shown in the figure by the dashed line.
Superposed on this general trend are persistent local maxima at $\sim$ 8, 11, and 14 au containing ``excess'' masses of $\sim 6M_{\rm J}$, $18M_{\rm J}$, and $10M_{\rm J}$, respectively.  As will be seen, these radii correspond with notable physical, kinematic, and thermodynamic ring-like features in the disk. Figure~\ref{fig:sigmaplot}(b) displays the mass enclosed on cylindrical shells with thickness of one radial cell length ($\sim 0.1667$ au) as a function of radius at this same time for the 512 simulation. Broad bumps are visible in the mass distribution centered around $\sim$31-32 au and 48 au.  These features fluctuate with time but are persistent. The location of the 31-32 au bump corresponds with a strong local maximum in time-averaged gravitational torques (Section 3.5). The 48 au bump arises from a strong one-arm spiral in the outer disk.  Masses interior to 10, 20, 30, 40, and 50 au are approximately 5\%, 20\%, 40\%, 62\%, and 86\% of the disk mass, respectively.  

\begin{figure}[htb!]
\hspace{0.5in}
\includegraphics[width=0.9\textwidth]{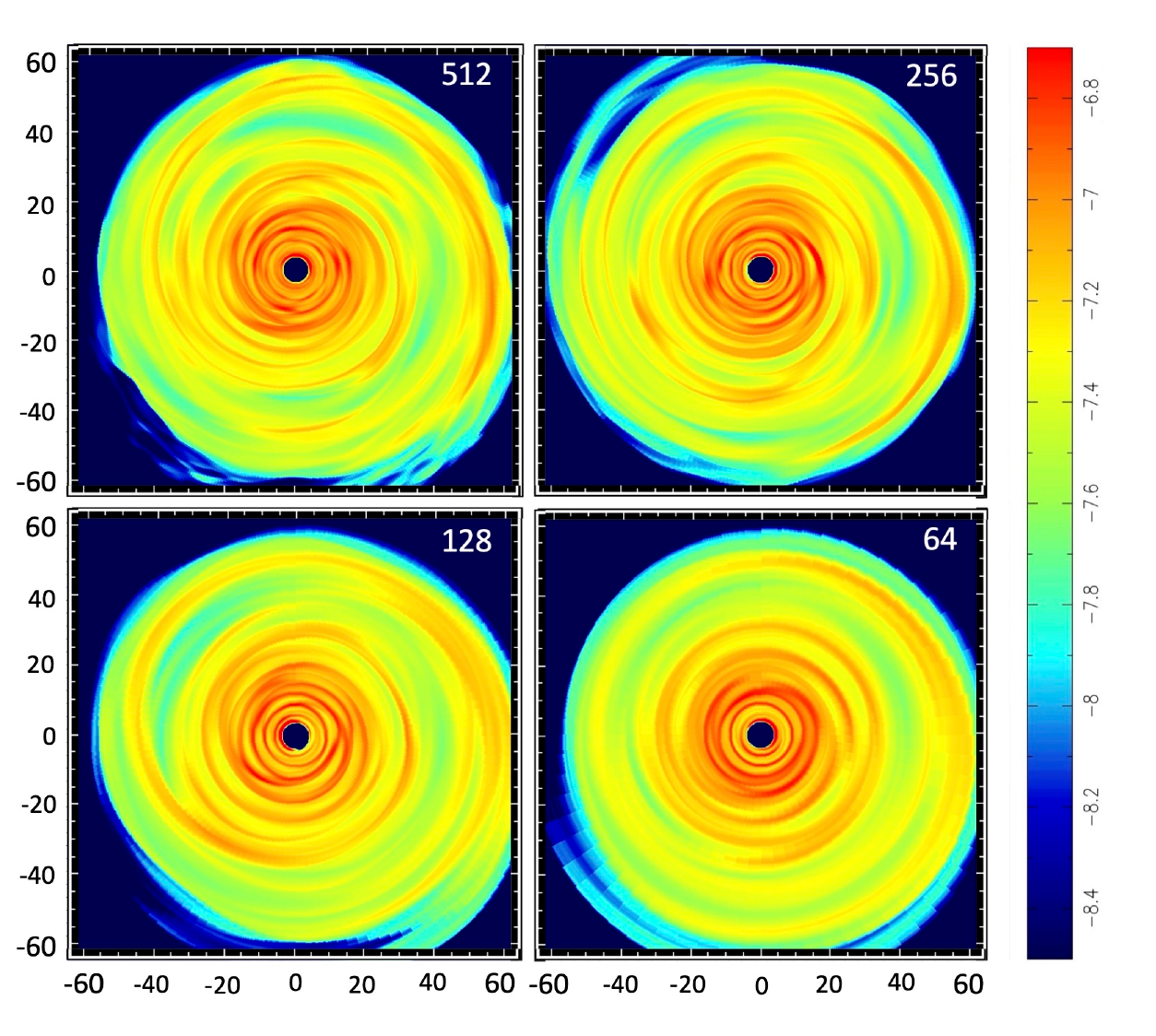}
\caption{
Enhanced density structures at $t \approx 20$ ORPs for the four simulations.  Panels are 120 au on a side and labeled with $l_{\rm max}$, the number of azimuthal grid elements used in the simulation.  Differences between the local surface density, $\Sigma(r,\phi)$, and the exponential fit to the azimuthally averaged surface density of Figure~\ref{fig:sigmaplot}(a), $\Sigma_{fit}(r)$, are represented by color contours.   The color scale is logarithmic in code units,  and axis units denote au. 
}
\label{fig:spiral_features_panel}
\end{figure}

Detailed nonaxisymmetric density structures can readily be seen by subtracting the exponential fit of Figure~\ref{fig:sigmaplot}(a) (dashed line) from the full 2D surface density distribution.  The resultant enhanced spiral density structures are shown in Figure~\ref{fig:spiral_features_panel}  at $t = 20$ ORPs.  The structure is similar at all resolutions but, as expected, not precisely identical. With higher resolution, the fine structure of the ring region is more pronounced and spiral structures tend to be sharper and more pronounced as $l_{\rm max}$ increases. New structure emerges as the resolution increases, but by 256 the global structure is roughly consistent. 


\begin{figure}[htb!]
\vspace{-10pt}
\centerline{\includegraphics[width=0.7\textwidth]{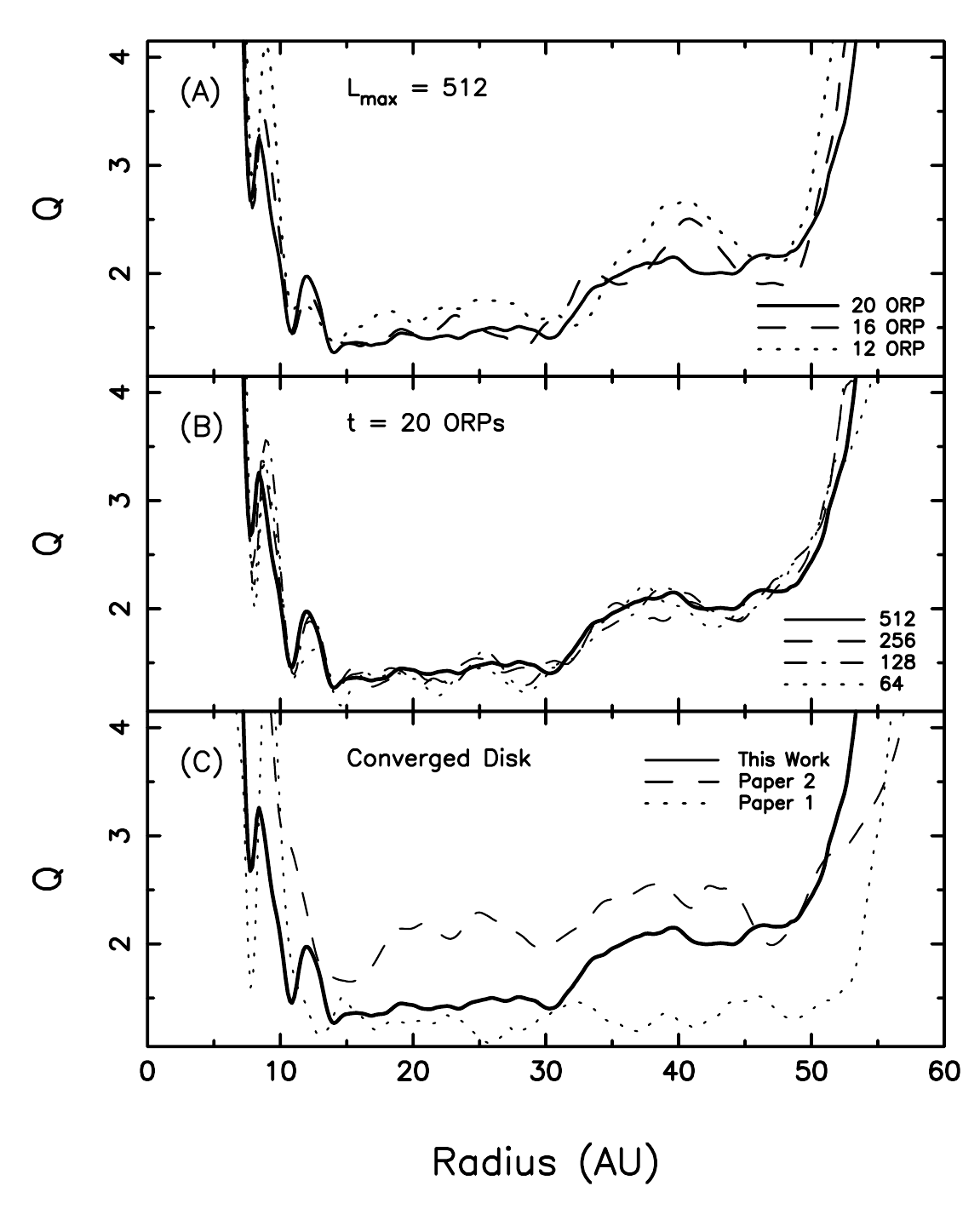}}
\caption{
(a) Azimuthally averaged $Q(r)$ for the $l_{max} = 512$ simulation, as a function of radius at 12, 16, and 20 ORPs.  By 16 ORPs, the disk has achieved a quasi-steady $Q$ interior to $\sim$ 40 au and is near that state out to the disk's outer edge. By 20 ORPs, the disk has settled over its entire radial extent.  (b) Azimuthally averaged $Q(r)$ for the four simulations, at $t =$ 20 ORPs.  By this time, all simulations have converged toward the settled 512 $Q(r)$ profile. (c)   Azimuthally averaged $Q(r)$ for the converged disks of Paper I (constant cooling), Paper II (no star--disk interaction, limiters on heating/cooling times), and the work reported here. }
\label{fig:qplot}
\end{figure}

A direct means of ascertaining convergence is provided by the characteristics of $Q(r)$ with time and azimuthal resolution.  A GI-active disk in its asymptotic state should be characterized by a quasi-steady $Q(r)$.  Figure \ref{fig:qplot}(a) shows azimuthally averaged $Q(r)$ for the $l_{\rm max} =  512$ simulation at $t =$12, 16, and 20 ORPs. Details describing how $Q$ was evaluated are described in detail in Section 3.1 of Paper II.

The inner disk relaxes faster than the outer disk.  At 12 ORPs, the 512 disk is still relaxing toward lower $Q$ over most of its radial extent.  By 16 ORPs, the disk has settled into an asymptotic state to $\sim$ 40 au.  We find that the entire 512 disk has reached this state by $\sim17$ ORPs (not shown in the figure) as represented by $Q(r)$ at 20 ORPs.   
At a given radius, $Q$ displays $\sim$5--25\% variability on the local dynamical timescale, with the amplitude of variability larger at larger radii. This was previously noted in Paper II (see Figure 4 of that paper). 

The middle panel of Figure \ref{fig:qplot} shows $Q(r)$ at 20 ORPs.  By this time, all four simulations have relaxed into similar, but not identical, asymptotic $Q(r)$ over the entire disk with clear convergence toward the $l_{\rm max} = 512$ curve. In what follows, we consider the state of the 512 simulation at $t=20$ ORPs as representing the quasi-steady asymptotic state for the system studied here, and we will refer to it as the ACDC, as defined at the beginning of Section 3.

As seen in the top two panels of Figure \ref{fig:qplot}, two radial regions exist in the ACDC within which $Q$ is essentially constant. {\it Region 1}, where $Q \approx 1.4$, lies between $\sim$11 and 32 au while {\it Region 2}, characterized by $Q \approx 2.1$, lies between $\sim$ 40 and 50 au.  Between these regions, $Q$ increases in a roughly linear fashion with radius.  A local enhancement in $Q$ between 11 and 14 au in Region 1 will be discussed in detail later.  $Q =$ 1.4 at the inner and outer edges of this feature, consistent with $Q$ in Region 1. Between the two regions, $Q(r)$ increases in a roughly linear fashion from $\sim 1.4$ to $\sim 2.1$.  Beyond 50  au, $Q$ rises sharply near the disk's outer edge as $\Sigma(r)$ goes to zero.


\begin{figure}[htb!]
\centerline{\includegraphics[width=0.9\textwidth]{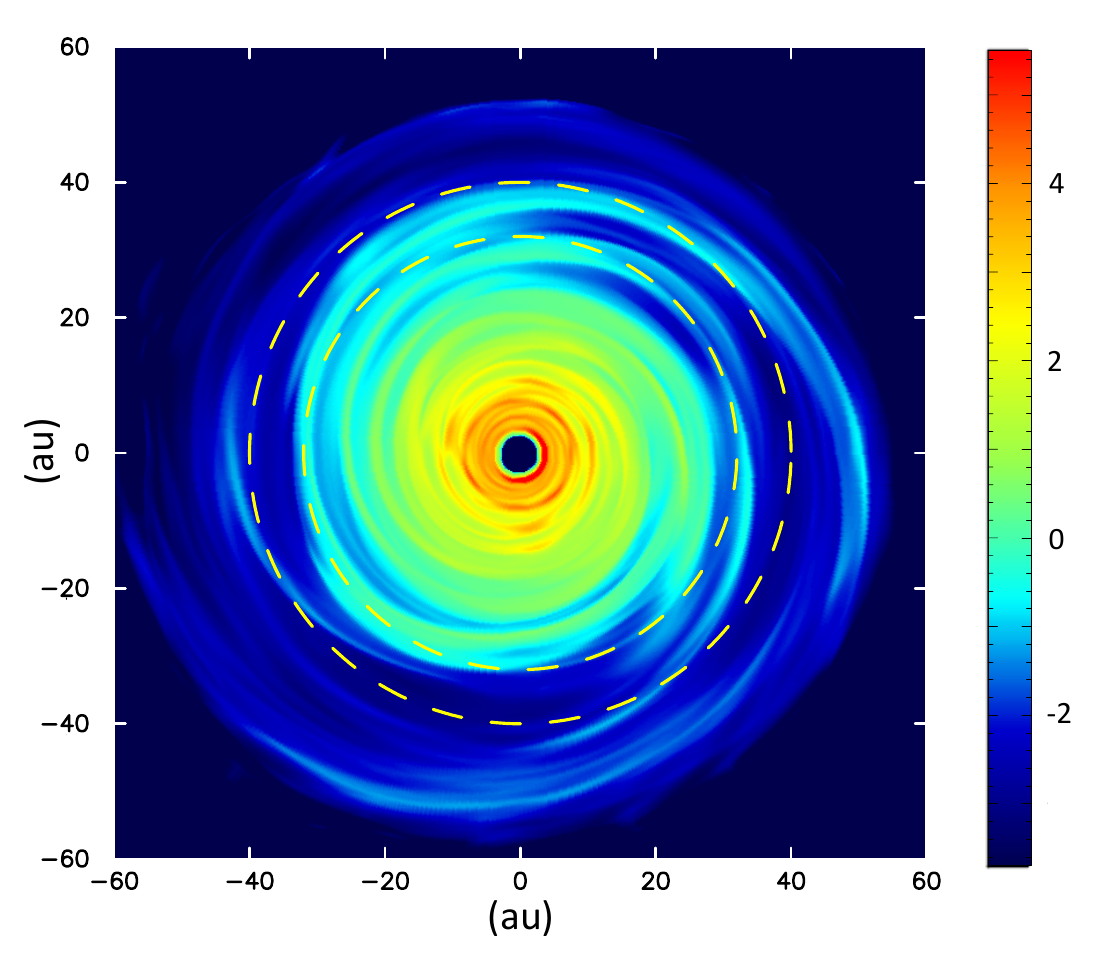}}
\caption{Optical depths normal to the disk plane of the 512 disk at $t=20$ ORPs. Colors show $\log(\tau)$.  Dashed circles at 32 and 40 au delineate the outer and inner radii of $Q$-defined Regions 1 and 2, respectively. }
\label{fig:2Dtauplot}
\end{figure}  


\begin{figure}[htb!]
\centerline{\includegraphics[width=0.7\textwidth]{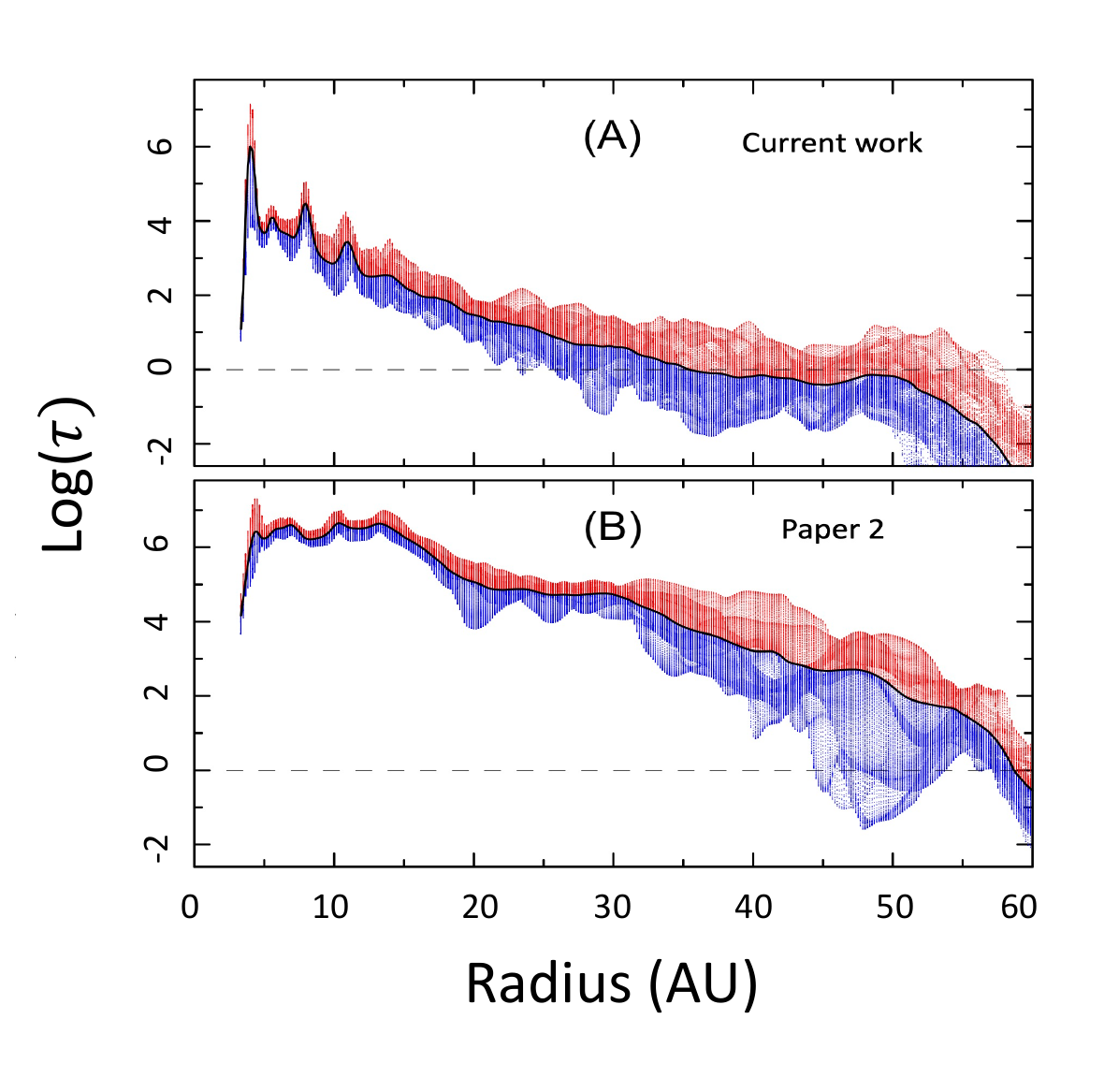}}
\caption{Optical depths normal to the disk plane at $t=20$ ORPs for all radial and azimuthal grid centers in the disk midplane.  Solid black curves trace the azimuthally averaged optical depth as a function of radius, red pixels correspond with grid cells with larger than the average at that radius, and blue pixels have smaller than the average at that radius.  Panel (a) shows the 512 disk of this work, and panel (b) shows the 512 disk of Paper II.  Note that the optical depths shown in Figure 9 of Paper II are in error.  Panel (b) has the corrected information and replaces that figure.}
\label{fig:tauplot}
\end{figure}  

Figure~\ref{fig:2Dtauplot} shows face-on color-coded optical depths of the 512 disk at $t=20$ ORPs with the inner and outer $Q$-defined Regions 1 and 2 depicted. 
The optical depth distribution is also shown in Figure~\ref{fig:tauplot}, which presents azimuthally averaged optical depths (solid black line) for the ACDC disk (panel (a)) and the analogous disk of Paper II (panel (b)).  Pixel representations in Figure~\ref{fig:tauplot} of individual azimuthal values of optical depth that are included in the average are displayed as blue pixels (less than average) and red pixels (larger than average).   Inspections of these two figures show that Region 1 in the ACDC disk is essentially optically thick while Region 2 is optically thin.  Between these regions the optical depth is a roughly equal mix of optically thick and thin cells.

The bottom panel of Figure \ref{fig:qplot} compares $Q(r)$ for the ACDC with the corresponding states of Papers I and II.  Interior to 30 au (Region 1), the ACDC $Q(r)$ looks very similar to the constant cooling ($t_{\rm cool} = 2$ ORPs) disk of Paper I but markedly different from the results of Paper II.  Paper II had larger asymptotic $Q(r)$ overall but the outermost disk and no extended regions where $Q(r)$ is flat.  Implementing a realistic, self-consistent, cooling approach in Paper II led $Q(r)$ to increase by $\sim 0.5$-- 1 relative to the values in the constant cooling simulations of Paper I. However, as described below, this was not due to the cooling approach itself, but rather to heating and cooling limiters implemented to maintain numerical stability.   

In Paper II, only in regions interior to $\sim 18$ au did the four simulations with differing $l_{\rm max}$ settle to approximately the same $Q(r)$ by 20 ORPs and the outer disk of the 512 model may not have fully relaxed by 20 ORPs.  We also note that Figure 9 of Paper II, which presents the same information as shown in Figure \ref{fig:tauplot}(b), is wrong owing to an integration error in the program that produced the Paper II figure.  This error has been corrected in \ref{fig:tauplot}(b) and it replaces the errant figure of Paper II.  The plotting error led to a misinterpretation of optical depths in Paper II.  In fact, virtually the entire disk of Paper II is optically thick, as shown in Figure~\ref{fig:tauplot}.  The plotting error should have had no effect on the hydro calculation of that work.  However,
as discussed in the next section, the use of limiters on cooling rates appears to have had adverse consequences on the outcomes of Paper II.

\subsection{Cooling Times and Temperatures} \label{subsection:heat_cool}

Thermal radiative cooling times $t_{\rm cool}$  are calculated on each cylindrical shell using an 
averaging-like scheme.  Specifically, the cooling time for the $j{\rm th}$ shell is given by
\begin{equation} \label{eq:tcool}
t_{cool} = \frac{\Sigma_{kl} \epsilon_{jkl}}{ \Sigma_{kl} \nabla\cdot F_{jkl}},
\end{equation}
which is the total internal energy $\epsilon$ in a cylindrical shell divided by the radiative energy-loss rate 
 in that shell. 
The $\nabla \cdot F$ terms  represent 3D radiative transport, as calculated in the code, but due to the cylindrical averaging, each $t_{\rm cool}$ is based on vertical and radiative energy transport only. 

The cooling time is useful for characterizing the disk, but it is not used directly by the code in any of the radiative gravito-hydrodynamic calculations. Rather, we use $t_{\rm cool}$ in this work primarily to compare the disk evolution here with the prescription used by \citet{gammie2001} discussed in Section 3.7, which uses a simple cooling time to parameterize energy loss. 

\begin{figure}[hbt!]
\centerline{\includegraphics[width=0.6\textwidth]{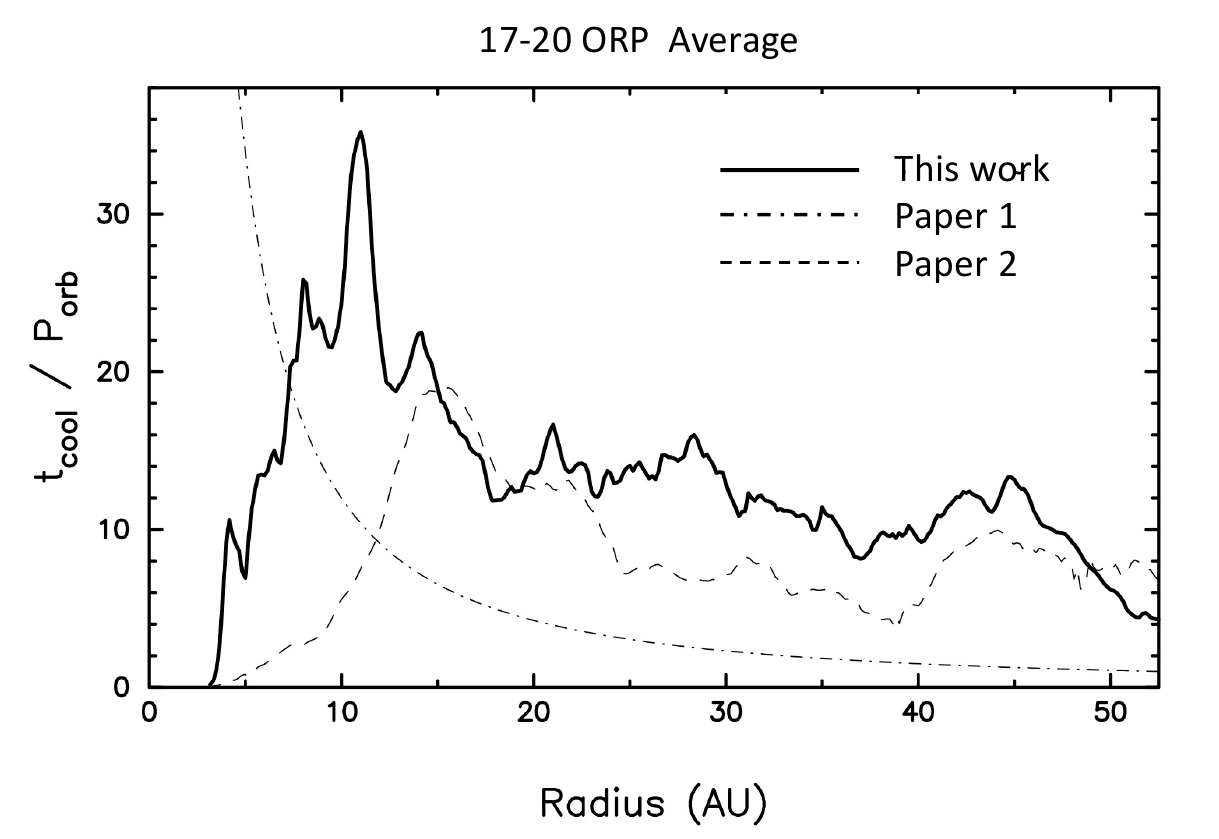}}
\caption{Azimuthally averaged cooling times normalized to the local orbital time for the ACDC disk of this work, Paper I, and Paper II.  Cooling times are averaged between 17 and  20 ORPS. }
\label{fig:Cool_Times}
\end{figure}

The resulting normalized cooling times for the ACDC disk are shown in Figure~\ref{fig:Cool_Times}, i.e., the cooling times given above, normalized to the local orbital period, $P_{\rm orb}$. In this case, the normalized cooling times are also time-averaged over 17--20 ORPs. For comparison, we also show the prescribed cooling times used in Paper I (constant cooling $t_{\rm cool}=2$ ORPs, no star--disk interaction) and the calculated cooling times found in Paper II (limiters on cooling rates, no star--disk interaction).


\begin{figure}[hbt!]
\centerline{\includegraphics[width=0.8\textwidth]{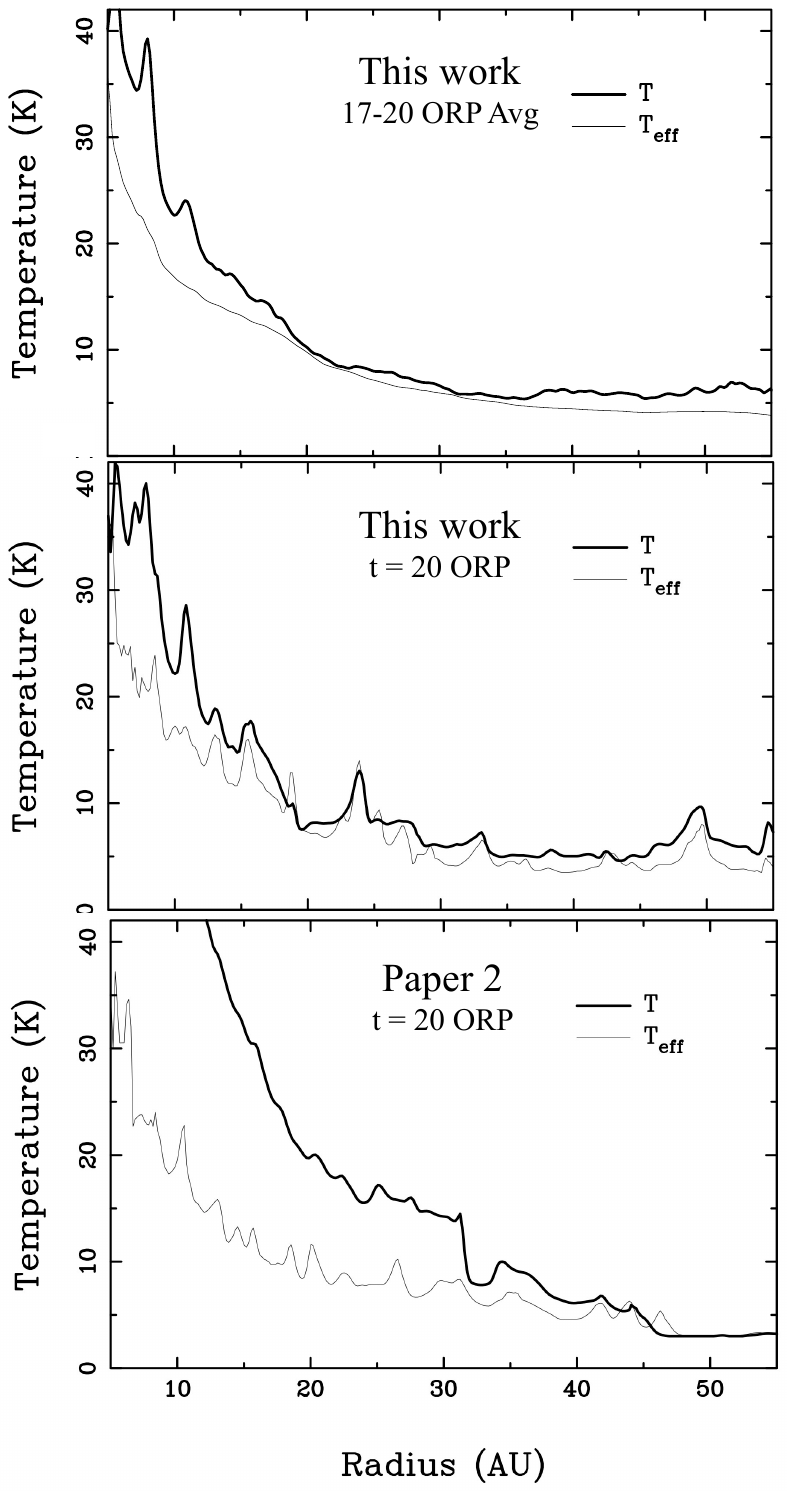}}
\caption{Azimuthally averaged midplane and effective temperatures  for the 512 disk of this work time-averaged over 17--20 ORPs (top), the 512 disk of this work at $t = 20$ ORPs (middle), and the 512 disk of Paper II at $t = 20$ ORPs (bottom).  We do not understand the ``glitch'' in the bottom panel that shows temperatures from Paper II, but note that this occurs at the same radius where the spread of optical depth about the mean at that radius undergoes an abrupt transition (see Figure \ref{fig:tauplot}).
}
\label{fig:Temperatures}
\end{figure}


\begin{figure}[hbt!]
\centerline{\includegraphics[width=0.75\textwidth]{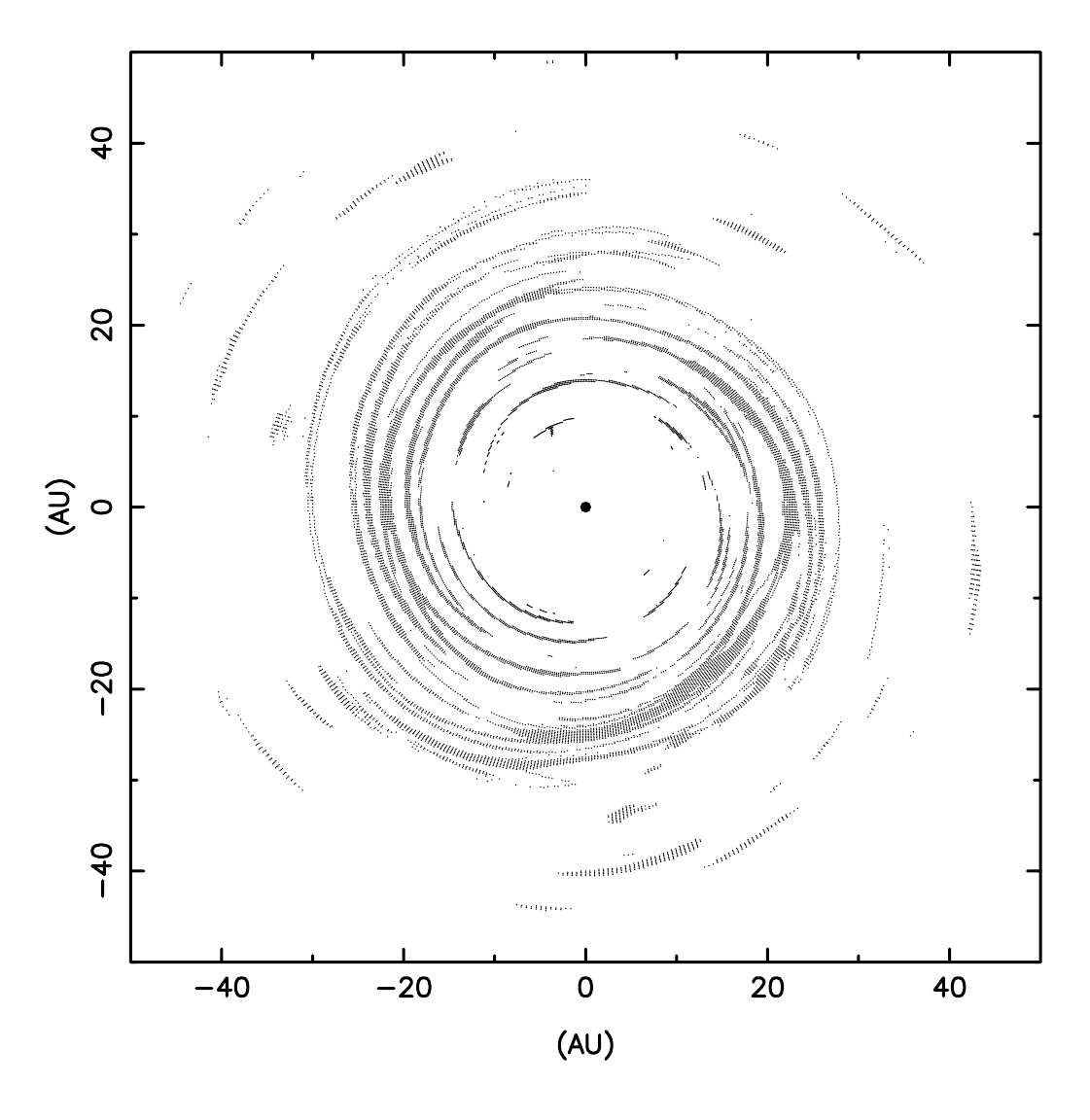}}
\caption{Locations of cells with negative divergent fluxes $\nabla\cdot F$ at 20 ORPs. These 
are strongly correlated with the locations of strong spiral waves seen in Figure~\ref{fig:spiral_features_panel}. 
}
\label{fig:Negative_divflux}
\end{figure}

In the ACDC disk, the normalized cooling times rise rapidly in the inner disk, peaking at 33 around 11 au. Due to known uncertainties in the calculations interior to 8 au (see Section 2.2), the cooling times inside this region should be interpreted with caution. At radii larger than 11 au, the normalized cooling times fall steadily, with an approximately linear decrease between about 15 and 40 au. These radii include the optically thick Region 1 and most of the transition to the optically thin Region 2. 

One might worry that the cooling times in the outer disk are problematically long. However, the divergence of the radiative flux includes the 3 K heating from the background. Normally, this is not relevant for the 
$t_{\rm cool}$ calculations because the effective temperatures are usually much higher than this. But in Region 2, the average effective temperature of the disk drops below 5 K and approaches 3 K.  Note that the ``effective temperature'' used here is really a brightness temperature determined from the outgoing radiative intensity in the vertical direction. It is only a measure of the radiation field, and not directly used in the simulation. It is nonetheless useful for understanding the behavior of $t_{\rm cool}$. In short, the temperature of the background radiation field contributes nontrivially to $t_{\rm cool}$ in the outer disk. With this, a long $t_{\rm cool}$ does not necessarily mean that the disk will require an equally long time to reach that state.

We can develop a more complete picture of the heating and cooling in the outer disk by instead looking at temperature and shock structures. Figure~\ref{fig:Temperatures} shows the radial, azimuthally averaged temperature in the midplane of the disk for a snapshot at 20 ORPs and for a time average between 17 and 20 ORPs. There are noticeable spikes in temperature at 20 ORPs that are not seen in the average, because such temperature spikes cool relatively quickly. The figure also shows that the effective temperature exhibits the same behavior. Any given snapshot can have sudden variations due to the spiral arms, but those temperature variations do not persist away from the shocks. For further context, the 3 K background is met at 58.5 au.

For comparison, the bottom panel of Figure~\ref{fig:Temperatures} shows the results from Paper II, which highlights that the ACDC disk is cooler over most of its radial extent. Because the surface densities are very similar between the ACDC and Paper II disks, the hotter temperatures in the Paper II disk explain the higher $Q$ values shown in Figure~\ref{fig:qplot}. We suspect this difference is primarily from the use of limiters, as already discussed in section 2.2.

One final point regarding the cooling times is that approximately 20\% of the computation cells at any moment have a negative $\nabla \cdot F$, which represents heating for our sign convention. An example of the locations of these cells is shown in Figure~\ref{fig:Negative_divflux}. They are strongly correlated with the strong spiral waves seen in Figure 3 and occur more commonly in optically thicker portions of the disk.  Their sharpness further suggests that these regions are associated with shocks, which are expected to have nontrivial radiative transport in all directions.

\subsection{Nonaxisymmetric Structures }

As demonstrated in Figure \ref{fig:qplot}(b), the ACDC is subject to GIs between 10 and 50 au, a region encompassing $\sim$76\% of the disk’s total mass, although, based on their asymptotic $Q$ values, we expect that these instabilities will be manifested differently in Regions 1 and 2.  Although the region interior to 10 au is not unstable to GIs, it is still affected by nonaxisymmetric density structures generated by GIs in Region I.

\begin{figure}[htb!]
     \centerline{\includegraphics[width=0.6\textwidth]{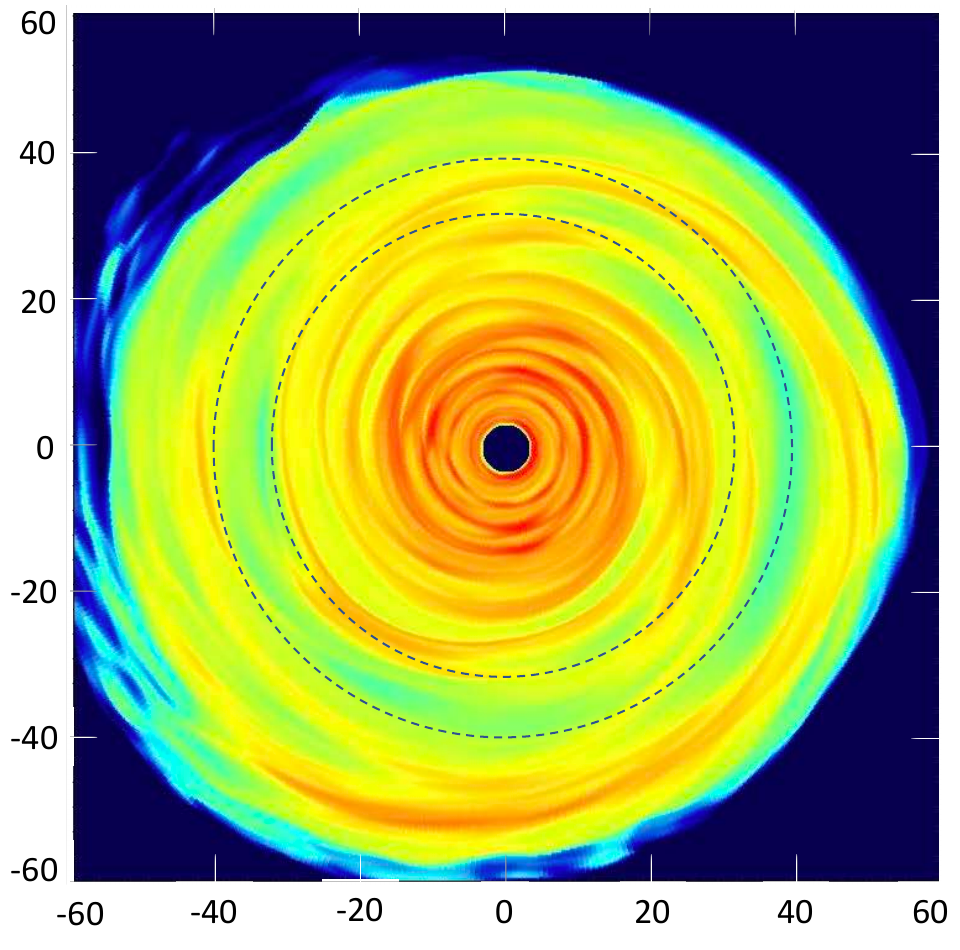}}
\caption{Enhanced density structures in the converged disk at 20 ORPs.  This figure is available as an animation showing the evolution of these structures between 17 and 20 ORPs.  Dashed circles are added at 32 and 40 au to delineate the outer and inner radii of $Q$-defined Regions 1 and 2, respectively.   Color contours show fractional differences between the local surface density, $\Sigma(r,\phi)$, and the linear fit to the azimuthally averaged surface density at that radius, $\Sigma_{fit}(r)$, shown by the dashed line in Figure \ref{fig:sigmaplot}(a).  The color scale is logarithmic in code units, and axis units denote au. }
\label{fig:FullDisk}
\end{figure}

Figure \ref{fig:FullDisk} shows the ACDC with spiral structures accentuated by plotting the difference between $\Sigma(r,\phi)$ and the exponential fit to azimuthally averaged $\Sigma(r)$ shown in Figure \ref{fig:sigmaplot}(a).  
Outer and inner boundaries of Regions 1 and 2, respectively, are delineated in the figure. 
An animation showing the time-evolution of Figure \ref{fig:FullDisk} can be accessed at 
\noindent
\begin{verbatim}
https://www.dropbox.com/s/7htomn38tb5vw0b/Fig9_Enhanced_Spirals.mp4?raw=1
\end{verbatim}
Visual inspection of Figure~\ref{fig:FullDisk} shows more complex spiral structures in the inner optically thick Region 1 than those seen in the optically thin Region 2. Indeed, structures with up to sixfold symmetry can be seen in Region 1 of the figure, while only a one-arm spiral is readily apparent in Region 2.


To quantify these nonaxisymmetric structures, we examine the Fourier amplitudes
\begin{equation} 
A_m=\frac{(a_m^2+b_m^2)^{1/2}}{\pi \int \rho_0 r drdz},
\label{eq:fourierameq}
\end{equation} where $\rho$ the mass density, $m$ is the order of the Fourier component, and $a_m$ and $b_m$ are the sine and cosine terms of the Fourier decomposition, respectively, as given by
\begin{subequations}
\begin{align}
a_m &= \int \rho \cos(m\phi)r {\rm d}r {\rm d}z {\rm d}\phi,\\
b_m &=\int \rho \sin(m\phi)r {\rm d}r {\rm d}z {\rm d}\phi.
\end{align}
\end{subequations}


\begin{figure}[htb!]
\vspace{-10pt}
\centerline{\includegraphics[width=1.0\textwidth]{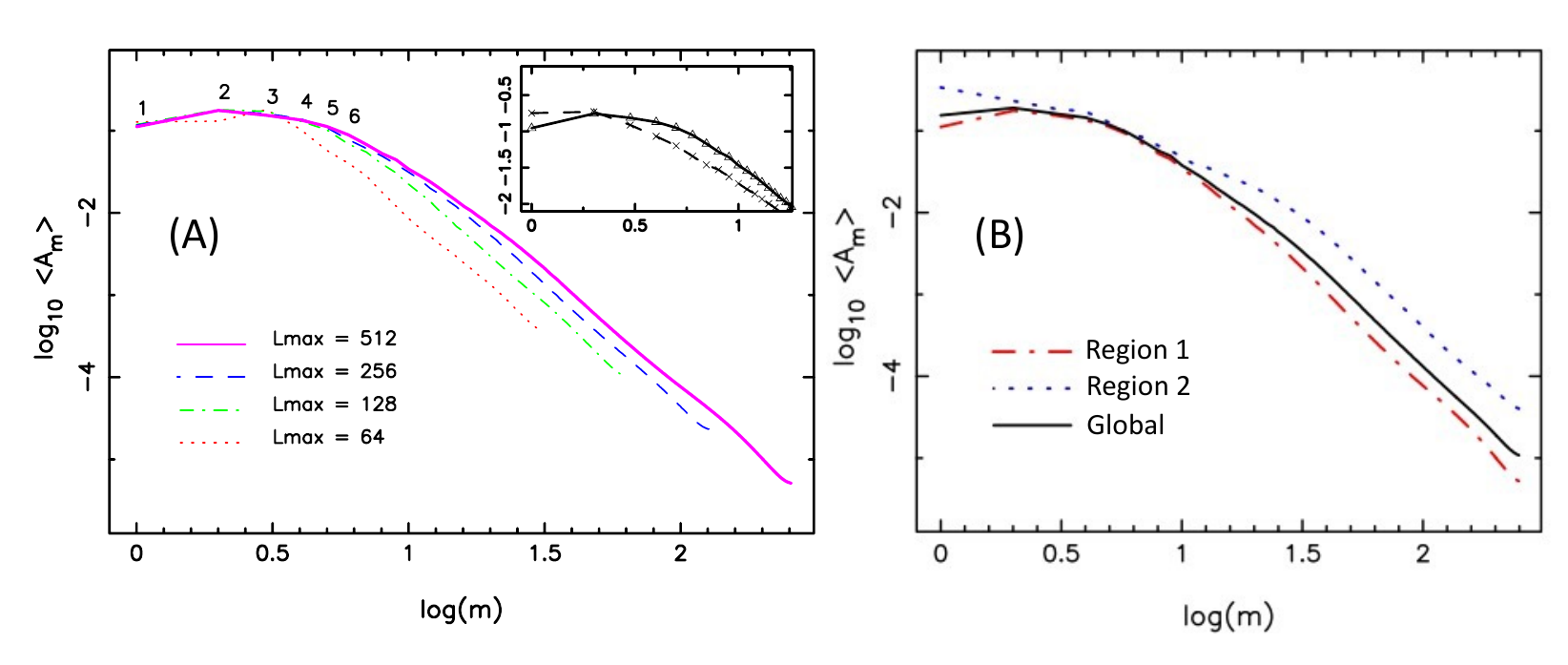}}
\caption{
(a) Strengths of the time-averaged global Fourier coefficients $\langle A_m \rangle$ integrated over 10--50 au for each azimuthal resolution.  Limiting resolutions are given by $m = l_{\rm max}/2$.  The inset compares $\langle A_m \rangle$ for the converged disk of this work (solid line with triangles) with the converged disk of Paper II (dashed with crosses).  (b) $\langle A_m \rangle$ calculated separately for Regions 1 and 2 in the converged disk along with the 512 $\langle A_m \rangle$ from panel (a).}
\label{fig:amplot}
\end{figure}
\noindent
The limiting azimuthal resolution for each $l_{\rm max}$ is given by $m = l_{\rm max}/2$.  GI-active disks display power at all resolvable $m$ values in their asymptotic phase \citep[e.g.,][]{lodato2004, mejia2005, boley2006, cossins2009, Michael2012,2016-ARAA-Kratter-Lodato}.    

For a disk-to-star mass ratio of the simulations presented here ($m_d/M_* = 0.14$),
dispersion studies predict that the power spectrum for a disk subject to quasi-steady GIs will be dominated by relatively loosely wound, low-$m$ waves \citep{LinShu1964, Vandervoort1970, LauBertin1978, Bertin2000, 2016-ARAA-Kratter-Lodato,Bethune_Latter_Kley2021}.

Figure \ref{fig:amplot}(a) shows global time-averaged Fourier amplitudes $\langle A_m \rangle$ for each simulation integrated over 10 -- 50 au and averaged from 16 to 20 ORPs.   As expected, the global power spectra are dominated by low-order Fourier components, and the spectrum continues to grow in amplitude at high $m$ as $l_{\rm max}$ increases. In this sense, while the spectra converge well by $l_{\rm max} = 512$, we cannot say that the spectra are converged. There is a residual uncertainty in our work in that we cannot be certain that, if the resolution were increased dramatically, we might actually see contributions from higher $m$ to the resultant transport.

The inset in Figure \ref{fig:amplot}(a) compares the lower-order $m$-terms $\langle A_m \rangle$ of the converged disk of this paper (solid line) with the corresponding converged disk of Paper II (dashed line).   Because star--disk interactions were not included in Paper II, the $m = 1$ component of the azimuthal mass distribution in that work was not accurately treated, leading to an artificially large $\langle A_1 \rangle$.  Indeed, the global power spectra of Paper II are dominated by the $m = 1$ and 2 Fourier components, much different than seen here.  This led to unrealistically large $m = $ 1 torques in the Paper II converged disk.  

Figure \ref{fig:amplot}(b) shows power spectra of the converged disk plotted separately for Regions 1 and 2 and the global results of panel (a).  
As expected from the visual appearance of Figure \ref{fig:FullDisk}, Region 2 is dominated by $m = 1$ with moderately strong $m = 2$.  Region 1 has strong contributions from $m = 1$ to 5.  For small asymptotic $Q$ (Region 1), many Fourier components, i.e., azimuthal symmetries, are important.  For larger asymptotic $Q$, only the lowest-order components appear dynamically important. 

From the perspective of numerical convergence, discussed in more detail in Section 3.9, the spectrum of amplitudes for high $m$ values in Figure~\ref{fig:amplot}(a) is not converging. However, the spectrum is converging for the lower-order $m$ values. As shown in Section 3.5, these dominate in the production of gravitational torques. 

We stress here that we have been examining time-averaged properties.  As will be demonstrated below, time averaging hides dynamically important time variabilities that affect the disk.

\subsection{Time Variability}

While the time-averaged Fourier analysis above gives insight into time-averaged, nonaxisymmetric disk structures, the radial and temporal stabilities of the Fourier components are important in determining their dynamical effects on the disk. For example, radial incoherence in a Fourier component will diminish its gravitational effects and lead the component to have more importance on a local scale than a global scale.  

\begin{figure}[htb!]
\vspace{-10pt}
\centerline{\includegraphics[width=1.0\textwidth]{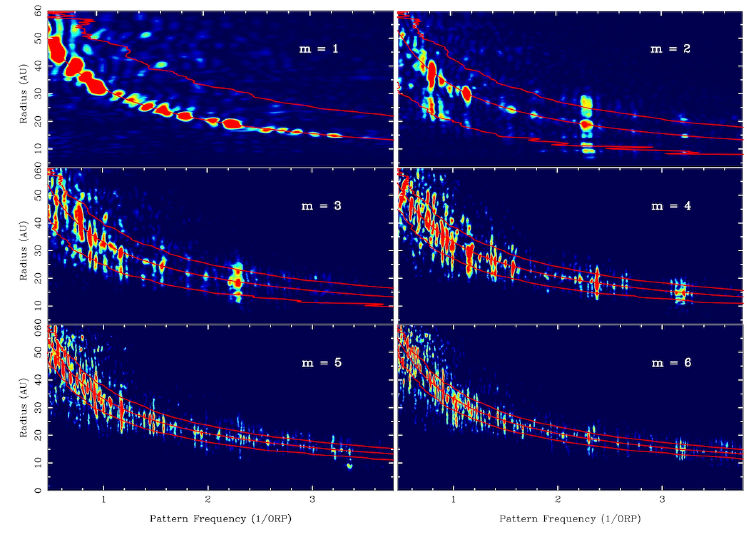}}
\caption{Periodograms for the $m = 1$--6 Fourier components of the converged disk during the 16--20 ORP time frame.  Red lines denote the corotation, inner Lindblad, and outer Lindblad resonances.  Zigzags in the ILR curves for $m=$ 2 and 3 are due to the non-Keplerian rotational frequencies, $\Omega(r)$, near the inner rings. 
}
\label{fig:PeriodogramPlot}
\end{figure}

Power in a specific $A_m$ does not imply the existence of an (eigen)mode for that $m$, nor does it necessarily represent the strength of an eigenmode that truly exists \citep[e.g.,][]{Michael2012,Steiman2013}. For example, a disk with a single $m = 2$ eigenmode growing to nonlinear amplitudes will exhibit power at all even values of $m$, while a disk with two nonlinear $m$ = 2 and 3 eigenmodes will exhibit power at all $m$ values. We will avoid referring to Fourier components as modes except for those cases where we have determined that a mode exists.  To this end, Figure~\ref{fig:PeriodogramPlot} displays periodograms \citep{scargle1982,horne1986,mejia2005, boley2006} of the converged disk for the  $m = 1$--6 Fourier components during the same 16--20 ORP time frame as Figure~\ref{fig:amplot}.  Locations of the corotation, inner Lindblad, and outer Lindblad resonances (CR, ILR, OLR) are displayed in each panel of Figure~\ref{fig:PeriodogramPlot}. Constructions of these periodograms use only the phase information $\phi_m$ from the Fourier decompositions of $\rho(r,\phi)$ in the midplane.   Power spectra of $\cos(\phi_m)$  are generated for each $r$-value using a large number of time steps over the time range 16--20 ORPs. If $\cos(\phi_m)$ is strictly periodic, i.e. $\phi_m$ is linear in $t$, then there will be a strong spike at the corresponding pattern frequency. 
These periodograms for all radii are combined into one plot in which isocontours of spectral power are traced.  If a pattern with a well-defined pattern frequency is present over a range of radii, it will produce a vertical stripe in the contour diagram.  Periodograms only measure the coherence, not 
the amplitude, of patterns present. Strong phase coherence combined with significant amplitude at the same $m$ value over the same radial range demonstrates that a dynamically significant $m$-armed wave is present.


The periodogram analysis of the 512 disk reveals a small number of unambiguous modes.  These include $m = 2$ modes at (ILR, CR, OLR) = (8, 15, 20 au), (11, 18.5, 25 au), and (22, 38, 48* au); an $m = 3$ mode with (ILR, CR, OLR) = (14, 18, 22.5 au); four $m = 4$ modes with (ILR, CR, OLR) = (11, 14.5, 17.5 au), (14, 17.5, 21 au), (21.5, 26, 32 au), and (28, 35, 43 au); an $m = 5$ mode with (ILR, CR, OLR) = (26, 32, 36 au); and an $m = 6$ mode with (ILR, CR, OLR) = (32, 36, 40 au). The m=2 mode with ILR, CR at (22, 38) would have an OLR around 48 au but does not have power in the periodogram extending to the OLR.  There are several other instances at all $m \ne 1$ with strong power between an ILR and CR radius that do not extend to the OLR.  In these cases, the OLR would fall in Region 2.

There are a number of well-defined, densely packed, high-$m$ stripes for CRs outside about 35 au and pattern frequencies $\sim$(1/ORP) or less.  These are ignored here owing to confusion and the fact that many have pattern periods roughly comparable with the 4 ORP time frame used in constructing the periodograms.

The strong ring-like structure at 8 au lies at the ILR of an $m = 2$ mode, the 11 au ring lies at the ILR of an $m = 2$ mode and the innermost $m = 4$ mode, and the 14 au ring is at the ILR of both $m = 3$ and $m = 4$ modes.  \citet{Durisen_etal_2005Icarus} previously noted similar ILR overlaps with positions of rings in their work on a hybrid theory of gas giant giant formation.

In addition to these well-defined modes, there are indications of transitory structures between the Lindblad resonances for a number of pattern frequencies and components.  For example, several of these can be seen as faint vertical strips in the $m = 4$ panel of Figure \ref{fig:PeriodogramPlot} at pattern frequencies between 1.4 and 2.6 ORP$^{-1}$ and spanning 10-30 au.  Similar features are visible for several $m$values.  These features represent density structures that have phase coherence for some period but less than the full time window of the periodogram.  One might think of them as modes that pop into and out of existence.  As will be seen, these {\it ephemeral modes} are dynamically important.  

The discussion of gravitational torques and mass transport in what follows (Sections 3.5 and 3.6) will make it clear that there are strong $m = 2$ and 3 effects for periods close to where the red blobs are in Figure~\ref{fig:PeriodogramPlot}. Hence, even if there are no pure sustained modes, the disk seems susceptible to bursts of global $m=2$ and 3 waves, probably swing amplified, and this is happening near pattern periods of about an ORP with CRs near 25-30 au. 

The ephemeral modes are consistent with the description of {\citet{Bethune_Latter_Kley2021} that gravitoturbulence generates spiral wakes that intermittently form and vanish over orbital timescales while, at the same time, large-scale spiral arms only manifest transiently through the coalescence of several neighboring wakes that are then are sheared apart.  However, as will be shown below, in addition to the ephemeral modes, we find recurrent (also on an orbital timescale) swing-amplified bursts that correspond with the modes listed above.  These bursts extend radially over the full ILR to OLR resonances.  In short, these are coherent modal structures.

\subsection{Mass and Angular Momentum Transport}\label{subsection:Transport}

Mass motions in accretion disks arise from stresses embodied in the stress tensor
\begin{equation} 
T = T^{\rm Rey} + T^{\rm grav} +  T^{\rm mag},
\label{eq:TorqueComponents}
\end{equation}
where 
$T^{\rm Rey}$, $T^{\rm grav}$, and $T^{\rm mag}$ represent the stresses arising from hydrodynamic (Reynolds), gravitational, and magnetic interactions.
Magnetic stresses fall outside the purview of this work and thus will not be considered further \citep[see][]{Deng_etal_2020ApJ, Bethune_Latter_2022AA}. 

The Reynolds stress tensor is defined by
\begin{equation}
T_{r\phi}^{\rm Rey}   \equiv  \frac{1}{2\pi r} \int_{A_{\rm cylinder}} \rho u_r' u_\phi' dA, 
\label{eq:newtonstress}
\end{equation}
where $\rho$ is the mass density and $u_r'$ and $u_\rho'$ are the fluctuations in the radial and azimuthal field components, respectively.
For the $i$-component of the velocity field, these fluctuations are  defined by $u_i' = u_i - \overline{u_i}$, where $u_i$ is the instantaneous velocity and $\overline{u_i}$ represents the ``mean'' (bulk) flow.  Unfortunately, it is not easy to properly determine Reynolds stresses in 3D nonlinear global simulations because of difficulties inherent in evaluating the local mean fluid flow.  In particular, the precise methods used to determine bulk flows are problematic.  In Paper I, we attempted to measure Reynolds stresses in similar 3D simulations using several different averaging schemes to evaluate the mean flow.  We found that different approaches yielded dramatically different results and there were no obvious criteria for selecting one approach over another. Several previous global 3D studies have reported that Reynolds stresses are small relative to gravitational stresses \citep[e.g,][]{lodato2004, boley2006, Michael2012, Steiman2013, Bae_etal2016ApJ, Bethune_Latter_2022AA}}. For these reasons, we omit Reynolds stresses in the calculations of angular momentum and mass transport.

The global torque, ${\rm \bf C}$, acting on a cylindrical section of the disk at radius $r$ can be calculated by integrating the stress tensor $T$ over the surface of the cylinder  \citep{lyndenbell1972}, i.e.,
 \begin{equation} 
{\rm \bf C} =\int \mathbf{r} \times T  dS.
\label{eq:torqueeq}
\end{equation} 
Since the stress tensor includes only gravitational stresses, the surface integral in Equation \ref{eq:torqueeq} can be replaced with the volume integral
\begin{equation}
{\rm \bf C}^{\rm grav} = \int \rho {\bf r} \times \nabla \Phi {\rm d}V,
\end{equation}
where $\Phi$ is the gravitational potential of the disk.  Here we are interested only in the $z$-component of torque,  
\begin{equation}
{\rm \bf C}^{\rm grav}_Z = \int \rho \frac{\partial\Phi} {\partial \phi} {\rm d}V,
\label{eq:cgravzeq}
\end{equation}
as only this component drives mass and angular momentum transport. 

The torque can be deconvolved into contributions from each
Fourier term by replacing $\rho$ in Equation \eqref{eq:cgravzeq} with the density distribution reconstructed from that Fourier component, i.e.,
\begin{equation}
\rho_m = a_{\phi m} \cos(m\phi) + b_{\phi m} \sin(m\phi),
\label{eq:rho_component}
\end{equation}
where $a_{\phi m} =\slantfrac{1}{\pi} \int \rho \cos(m\phi){\rm d}\phi$,  $b_{\phi m} =\slantfrac{1}{\pi} \int \rho \sin(m\phi){\rm d}\phi$, and only the gravitational potential produced by the mass distribution given by $\rho_m$ is included in $\Phi$. The total torque is then the sum of these torque components.


\begin{figure}[htb!]
\vspace{-10pt}
\centerline{\includegraphics[width=0.65\textwidth]{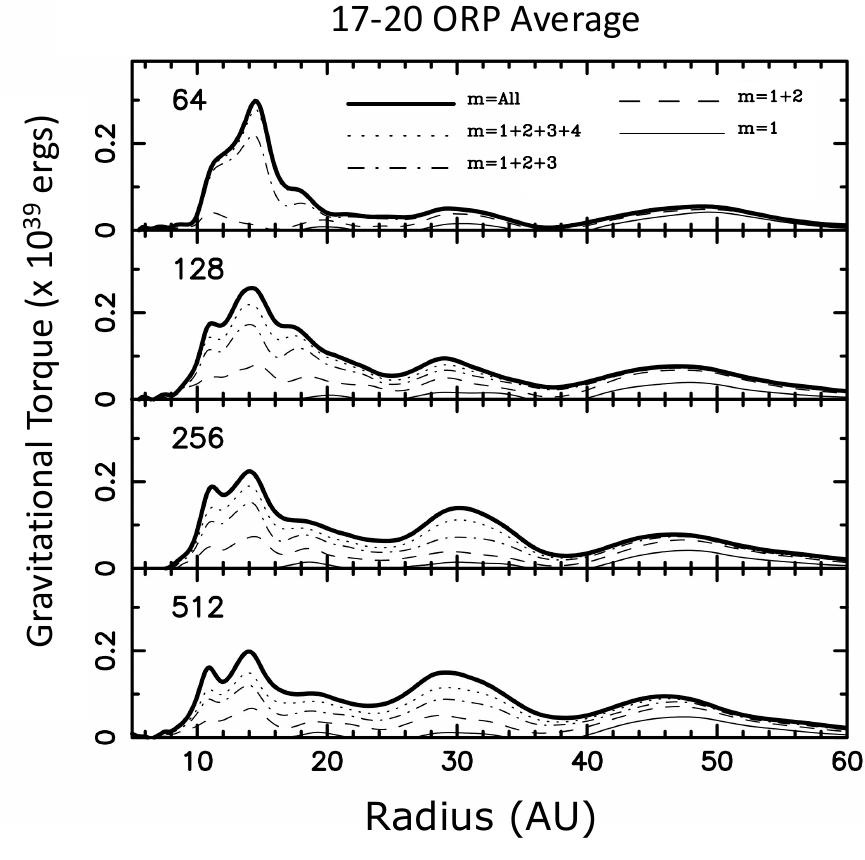}}
\caption{Time-averaged gravitational torques as a function of radius for each of the four simulations.  The total gravitational torque ($m = $ All curve) is shown along with contributions to the total arising from low-order ($m = $ 1 --  4) Fourier components of the azimuthal mass distribution.  These are shown as summations of the individual components.  Torques are averages of instantaneous torques at 240 equally time-spaced times between 17 and 20 ORPs.
}
\label{fig:torque_convergence}
\end{figure}

Figure~\ref{fig:torque_convergence} displays the total time-averaged gravitational torque, $\langle{\rm \bf C}_Z(tot)\rangle$, and time-averaged torques summed over a number of low-order Fourier components, $\sum_1^m \langle {\rm \bf C}_{Z(m)}\rangle$, $m =$ 1 -- 4, for each of the four simulations.  All torques are time-averaged over 240 equally spaced times between 17 and 20 ORPs to suppress short-timescale fluctuations. 

With the exception of the $l_{\rm max}=64$ simulation, time-averaged torques are dominated by several low-order ($m \sim 2$ -- 6) components throughout the optically thick Region 1 and much of the transition region, with $m =$ 2 -- 6 providing $> 95$\% of the total torque and no one component dominating the time average.  In the optically thin Region 2, $m =  1$ and 2 dominate.  Some $m=1$ strength may be due to beating of $m=2$ and $m=3$ 
but we suspect that most of its strength arises from sling amplification, a type of eccentric GI in nearly 
Keplerian disks \citep{Shu_etal_1990ApJ,Ostriker_Sling2_1992ApJ,2016-ARAA-Kratter-Lodato}.

These results, where several low-order Fourier components dominate the optically thick regions and low-order terms dominate optically thin regions, are consistent with several other studies (see Section 3.3).  The $l_{max}=64$ disk is distinctly different in that it is dominated by $m=3$ torques for $r \leq 20$ au, $m=2$ torques in the $r = $ 25 -- 35 au region, and $m=1$ torques outside 40 au.

Clear convergence toward the 512 disk is visible.  Torque profiles of the 64 and 128 simulations
show very clear differences with the 256 and 512 disks while the 256 and 512 torque profiles are essentially identical.


\begin{figure}[htb!]
\vspace{-10pt}
\centerline{\includegraphics[width=0.75\textwidth]{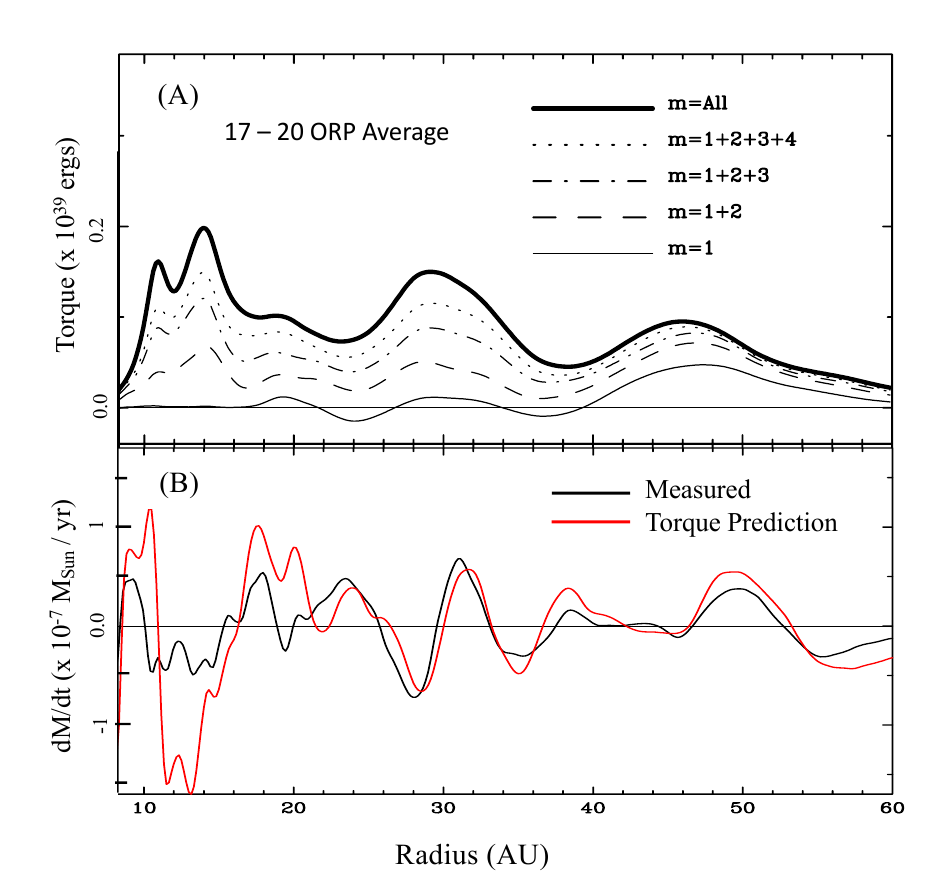}}
\caption{(a) Time-averaged gravitational torques of the converged disk, repeated here from the bottom panel of Figure~\ref{fig:torque_convergence}.  Plotted torques are averages of instantaneous torques at 240 equally time-spaced times between 17 and 20 ORPs.  (b) Measured and torque-predicted time-averaged mass fluxes. Measured fluxes are obtained directly by averaging the time variability of masses on cylindrical shells at the same times used in determining the gravitational torques on the top panel.  It is a direct output product of the simulation.  Predicted $\dot{M}(r)$ are determined by applying Equation~\ref{eq:Mdot_grav_torques} to the instantaneous total torques used in generating the time-averaged torques of panel (a) and then averaging the resultant instantaneous $\dot{M}(r)$.  
The disagreement interior to $\sim$ 22 au likely arises from unaccounted-for stresses.}
\label{fig:torqueplot}
\end{figure}

The results found here for the converged disk are similar to those found in the constant cooling study of Paper I but considerably different from those of Paper II.   In Paper II, a very strong $m =  2$ mode exists and is the dominant torque over most of the disk with significant, but not dominant, contributions from $m = 1$ in selected radial regions centered around 28 and 38 au.  We attribute most of the problematic results of Paper II to the use of cooling limiters in that study. The results in this paper using radiative subcycling supersede the results of Paper II.

Radial mass fluxes arising from the gravitational torques of Equation \ref{eq:cgravzeq} can be written as \citep[see also][Eq. 28]{balbus1999}
\begin{equation}
\dot{M} = \frac{2}{r\Omega} \frac{d}{dr} (C_z).
\label{eq:Mdot_grav_torques}
\end{equation}


\begin{figure}[htb!]
\vspace{-10pt}
\centerline{\includegraphics[width=1.0\textwidth]{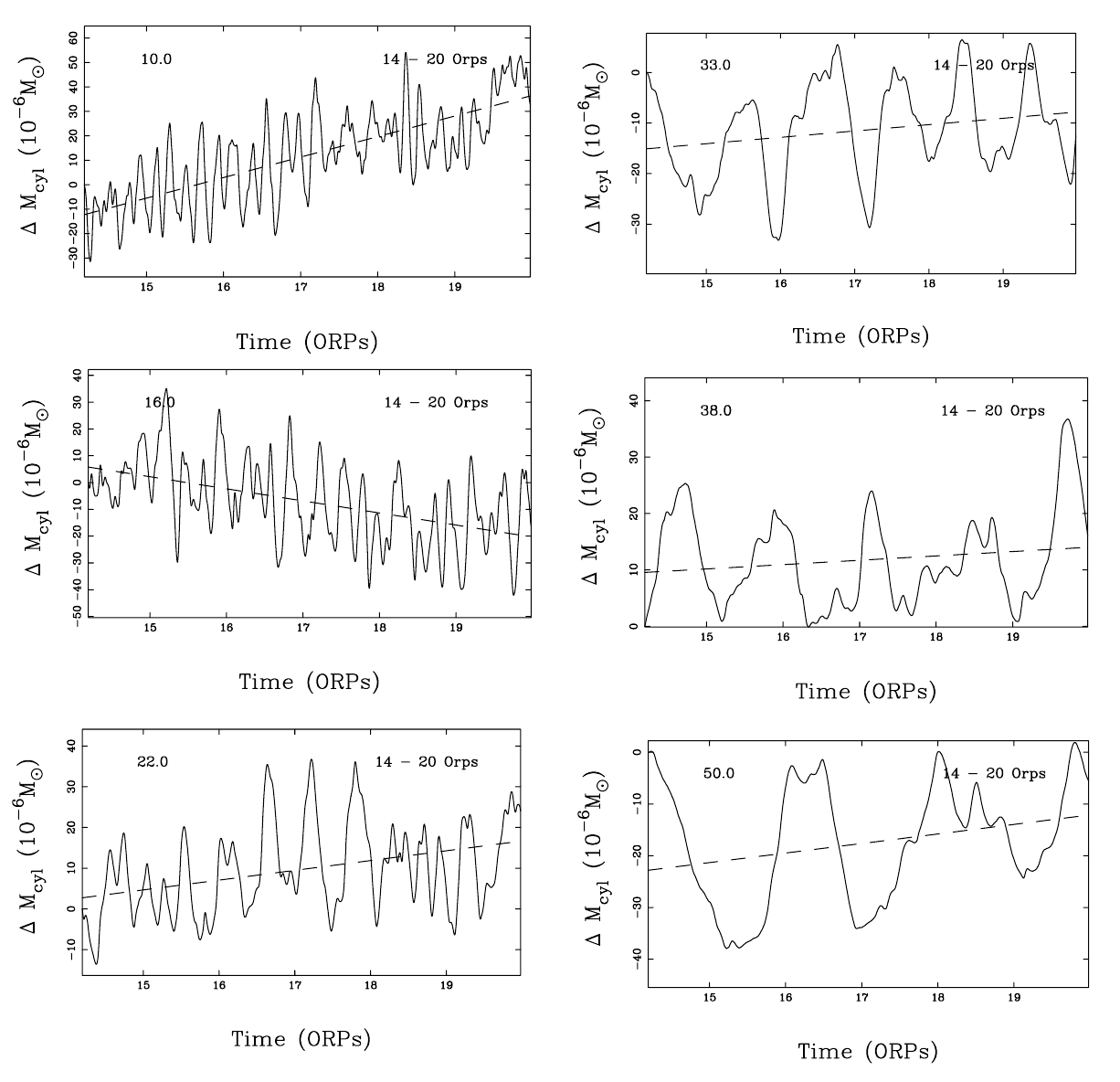}}
\caption{
Time variability of the mass in shells at six radii.  At each radius (labeled in the upper left corner of the panels), the solid line shows the change in the cylindrical mass, $\Delta M_{\rm cyl}(t)$, measured relative to the mass at 14 ORPs.  Cylindrical masses vary 
by $\sim2$--10\%  in a roughly cyclical manner on the local dynamical time. GI-active disks slosh around considerably more mass on short timescales than the net transport, which is indicated in the diagrams by the dashed linear fits.  This sloshing can be seen in the animation associated with Figure~\ref{fig:Mass_on_Cylinders}.
}
\label{fig:DeltaMass_fits}
\end{figure}


\begin{figure}[htb!]
\centerline{\includegraphics[width=1.0\textwidth]{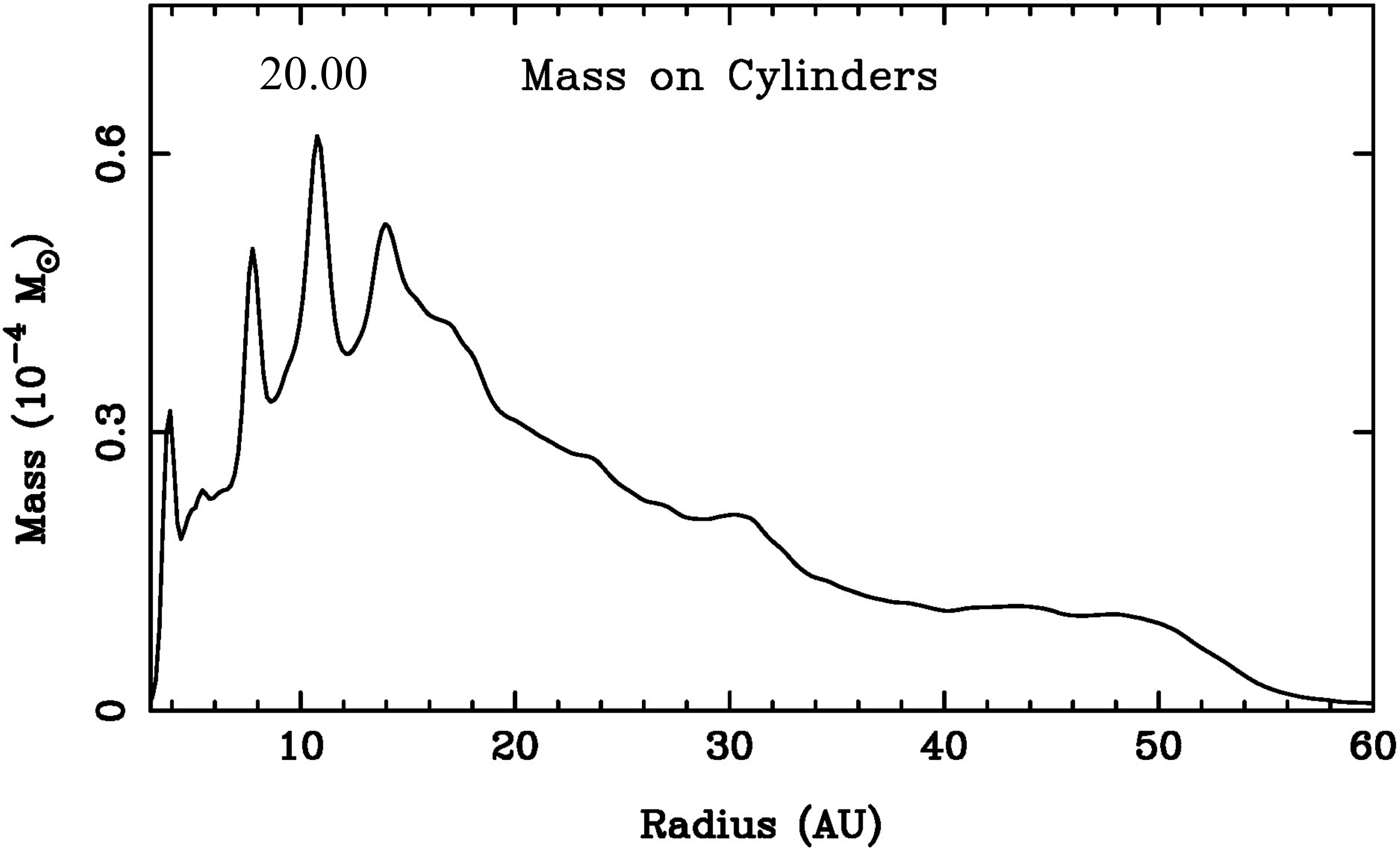}}

\caption{
Mass on cylinders.  This figure, available as an animation, shows masses in $0.167$ au wide cylindrical shells, $M_{\rm cyl}$.  Each frame is labeled with the time in ORPs. The animation runs from 17.00 to 19.96 ORPs in 0.03 intervals. The real-time duration of the animation is 8 s.
}
\label{fig:Mass_on_Cylinders}
\end{figure}

Figure~\ref{fig:torqueplot}(b) shows mass fluxes predicted by applying this equation to the time-averaged instantaneous torques used in generating Figure \ref{fig:torqueplot}(a).  These gravitational torque-predicted rates range between $\sim \pm 10^{-7} \sim M_\odot/{\rm yr}^{-1}$, rates comparable to accretion rates reported in other global 3D simulations. 
These radial mass flows correspond to evolutionary times $\sim10^6$ yr. It is important to note that there are several bands with different directions of radial flow at different ranges of $r$, very different from a simple $\alpha$-disk.

For comparison, the panel shows ``measured'' fluxes obtained directly from the output products by measuring how the mass within a cylindrical shell one radial grid element thick changes with time.  Instantaneous $\dot{M}(r)$ are determined at the same 240 times used in determining the gravitational torques in panel (a) and then time-averaged.

Predicted and measured mass fluxes agree well for radii larger than $\sim 22$ au but poorly at smaller radii. The disagreement interior to $\sim$ 22 au suggests that some stresses or transport terms are not properly accounted for, possibly Reynolds stresses.  

We discuss below in Section \ref{subsection:short_timescale_variability}, but note here, that the region in the inner disk where the red curve shows poor agreement is a region of strong torque outbursts and a region of strong shocks, which suggests strong systematic (not turbulent) flows in the spiral arms, which are Reynolds stresses when rotation is used as the ``mean flow.'' 

We also note that masses on shells significantly change, in both the positive and negative sense, on local dynamical time scales or less, as shown in Figure~\ref{fig:DeltaMass_fits}.  These short-timescale changes, in turn, follow longer-term trends.  These can make measured $dM/dt$ difficult to accurately assess. 
These variations are readily apparent in the animation of Figure~\ref{fig:Mass_on_Cylinders}, which shows how the radial mass distribution profile $M_{\rm cyl}(r)$ changes with time.  On timescales of the local dynamical time, the mass distribution in the disk displays radial oscillations giving the appearance that the disk “sloshes.'' 
This animation can also be seen at
\begin{verbatim}
https://www.dropbox.com/s/q4p1a1lrfc0kp44/Fig14-Mass_on_Cylinders_Animation.mp4?raw=1
\end{verbatim}


\begin{figure}[htb!]
\centerline{\includegraphics[width=1.0\textwidth]{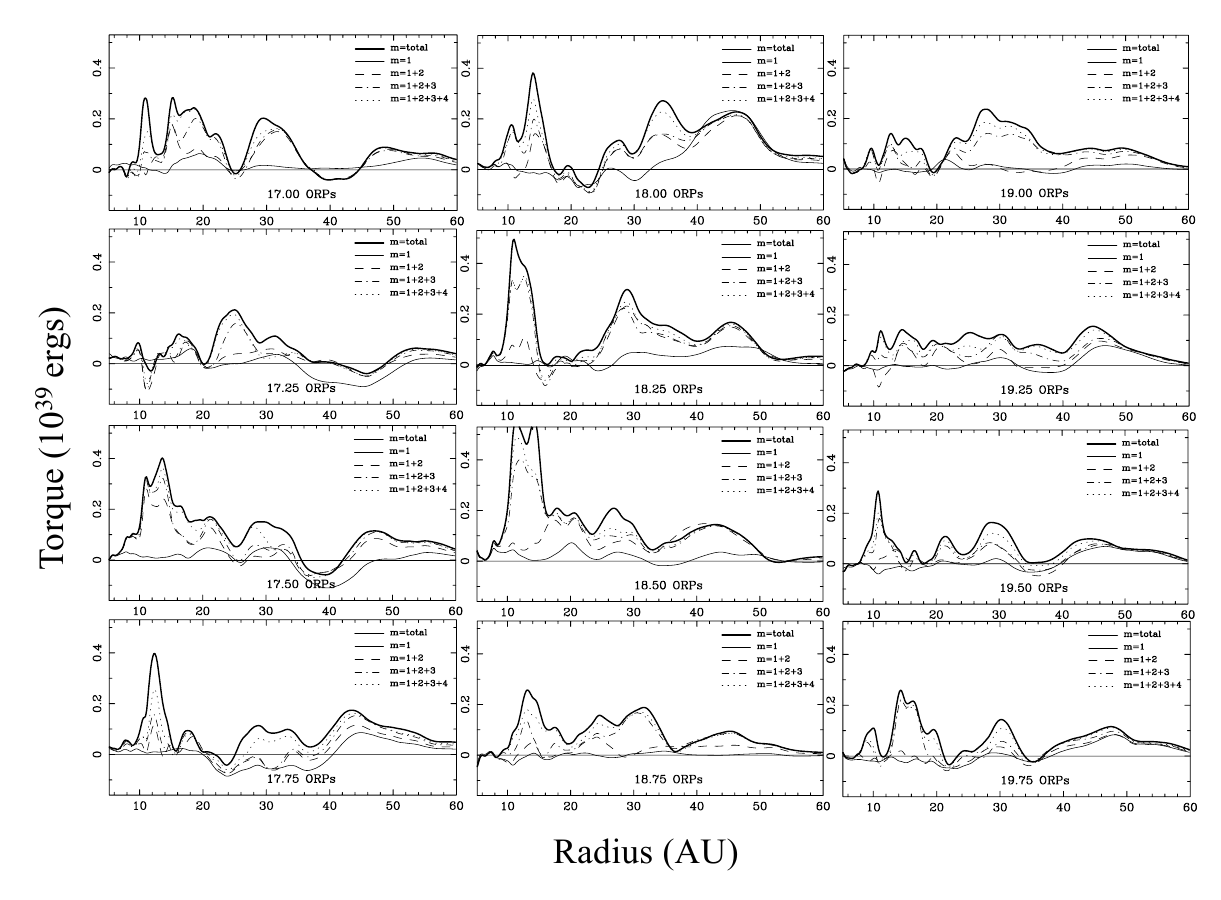}}
\caption{Instantaneous gravitational torques.  This figure is available as an animation.
Panels are labeled with the times, in ORPs, for which the torques are plotted. The elapsed time between panels (0.25 ORPs) is approximately 64 yr.  Total torques (the sum of all $m$-terms) and summations of low-$m$ torque components ($m = 1$, 1+2, 1+2+3, and 1+2+3+4) are depicted.  The key is the same as that used in Figure \ref{fig:torqueplot}(a).  As is clear from the figure, torque components and total torques are highly variable on short timescales. The animation runs from 17.00 to 19.93 ORPs in 0.012 ORP intervals. The real-time duration of the animation is 40 s.
}
\label{fig:variable_torques}
\end{figure}


\begin{figure}[htb!]
\vspace{-10pt}

\centerline{\includegraphics[width=0.88\textwidth]{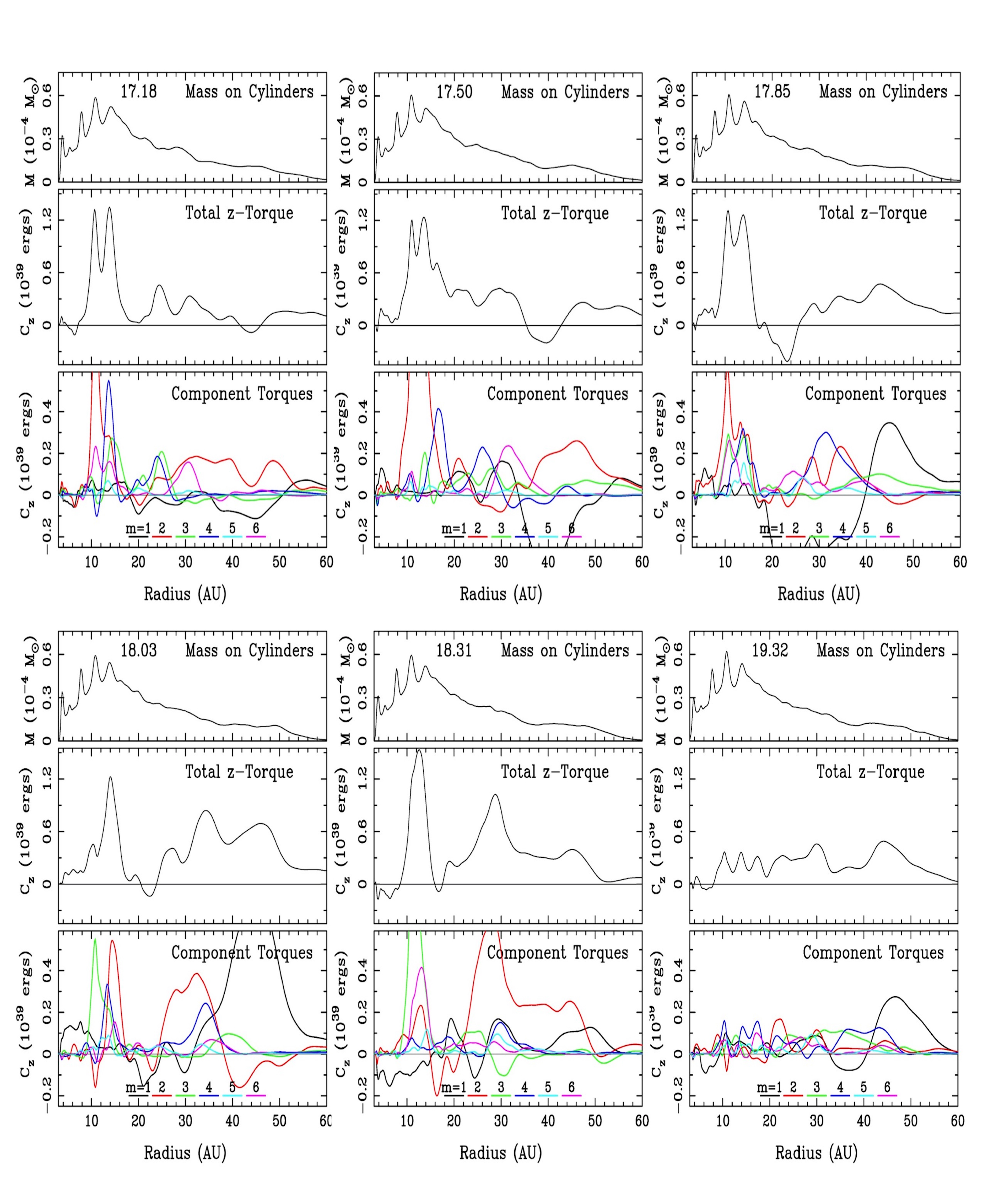}}

\caption{Total torques, low-$m$ torques, and masses on cylinders as a function of radius at six points in time.  This figure is available as an animation.  Note recurrent $m =$ 2, 3, and 4 amplitude bursts, which we attribute to swing amplification, and bursts of $m =1$ amplitude at larger radii, which we attribute to sling amplification.
Individual frames in the still figure are selected to display a wide range of different torque activity more readily visible in the animation.
The animation runs from 17.00 to 19.98 ORPs in 0.01 ORP intervals. Time is shown at the top of each panel. The real-time duration of the animation is 48 s.
}
\label{fig:BigPicture}
\end{figure}


\begin{figure}[htb!]
\vspace{-10pt}
\centerline{\includegraphics[width=1.0\textwidth]{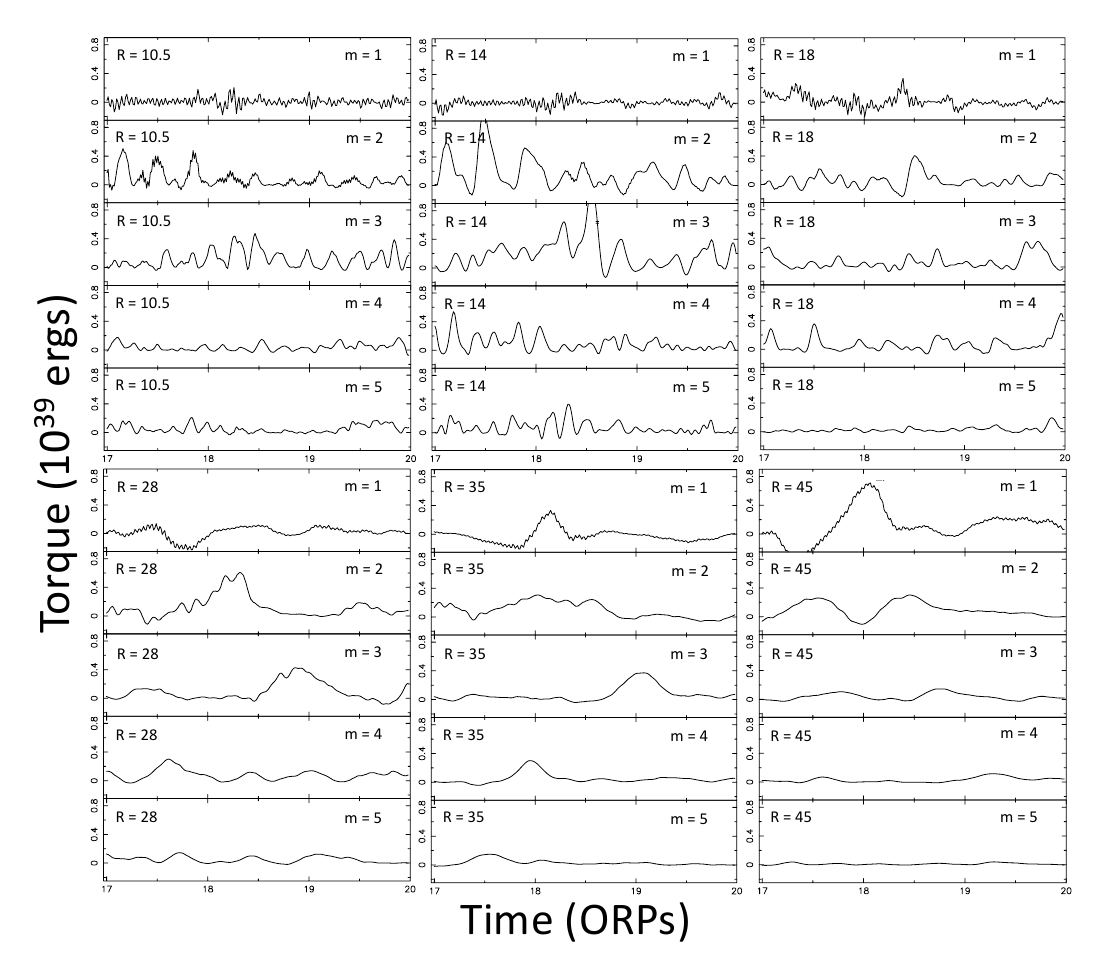}}
\caption{Strengths of low-order Fourier component torques at selected radii as a function of time.
Note recurrent amplifications of $m = 2$--5 torques at radii interior to $\sim$18 au.}
\label{fig:Torque_Frequencies}
\end{figure}


\begin{figure}[htb!]
\vspace{-10pt}
\centerline{\includegraphics[width=0.9\textwidth]{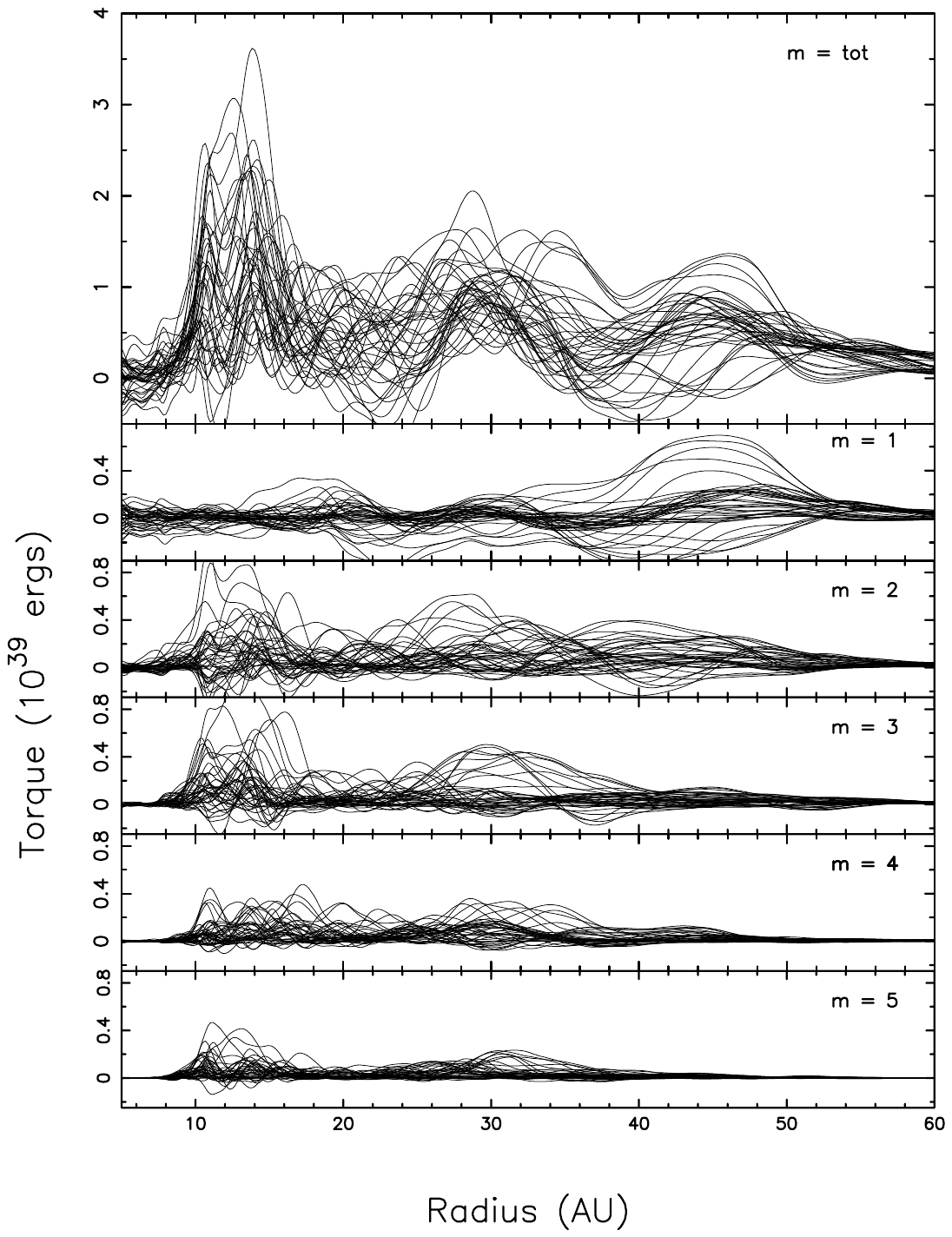}}
\caption{Overplotted instantaneous gravitational torques arising from $m =$ 1-5 Fourier components and the total torque at 41 equally spaced times between 17 and 20 ORPs.  This demonstrates how variable torques are on short timescales.  It also illustrates how there are strong bursts in the torques, especially for $m = $ 2 and 3 in Region 1 and $m = $ 1 and 2 in Region 2.
}
\label{fig:Torque_Range}
\end{figure}

\subsection{Short-timescale variability}\label{subsection:short_timescale_variability}

GI-induced mass transport arises from nonaxisymmetries in the mass distribution and hence nonaxisymmetries in the gravitational potential of the disk plus star.  These nonaxisymmetries change continuously with time, as seen by changes in the geometry and strengths of spiral structures in the animation associated with Figure \ref{fig:FullDisk}.  Fluctuations in strength and coherence manifest themselves as the ephemeral modes seen in the periodogram of Figure \ref{fig:PeriodogramPlot} leading to the expectation that gravitational torques and mass flows also display considerable variations on short timescales.  

Figure \ref{fig:variable_torques} shows instantaneous gravitational torques, including low-order Fourier components and total gravitational torques, at nine points in time between 18 and 20 ORPs, at intervals of 0.25 ORPs  ($\sim$ 64 yr) with low-order torque components labeled the same as in Figure \ref{fig:torqueplot}(a).  
An animation of Figure~\ref{fig:variable_torques} can be accessed at:
\begin{verbatim} 
https://www.dropbox.com/scl/fi/uj8l9nimwes3lvxbul55r/Fig17-Low_m_torques_NEW.mp4?dl=0
\end{verbatim}

Instantaneous torques vary dramatically on timescales less than the local dynamic time, often by 50\% or more, and sometimes exhibit brief changes in the sign of local torque.  Large local increases in the total torque are typically produced by sudden strengthening in an individual particular Fourier component, strongly suggesting recurrent bursts of swing amplification.

This can also be seen in Figure \ref{fig:BigPicture} which shows the instantaneous radial mass distribution, total $z$-torques, and torques from low-order Fourier components at nine points in time.  Individual time frames are selected to show the wide range of burst activity seen at different times.  Three vertical frames are associated with each point in time.  These show: (1) the mass enclosed within a cylindrical shell one radial cell width wide ($\sim$ 0.1667 au), (2) azimuthally averaged total torques, and (3) torque contributions from $m =$ 1, 2, ..., 6 Fourier components.  The bottom right panel ($t = $19.32 ORPs) displays a time with relatively small torques over the full disk interior to 40 au with no Fourier component dominating the total torque.  All other frames show strong bursts arising from one or more components, with individual component torques often changing sign.  

The animation associated with Figure \ref{fig:BigPicture} shows that, with the exception of $m=1$, bursts typically have durations on the order of the local rotation period as expected for swing amplification ($m=1$ is not subject to swing amplification).  These recurrent amplifications are readily apparent in Figure \ref{fig:Torque_Frequencies}, which shows the time-progression of torques associated with the $m = 1$ -- 5 components at six radii between 10 and 45 au.  
An animation showing the time progression of torques and torque components between 17 - 20 ORPs in the same format as Figure \ref{fig:BigPicture} is available and can also be found at
\begin{verbatim}
https://www.dropbox.com/scl/fi/vs5i1odswfxrga7b1oqxq/Fig18-Animation.mp4?dl=0
\end{verbatim}

Figure \ref{fig:Torque_Range} overplots the total torque and torque contributions from $m = 1$ -- 5 components at 41 equally spaced intervals of $\sim$ 25.5 yr between 17 and 20 ORPs.   Readily noticeable in Figures \ref{fig:variable_torques} - \ref{fig:Torque_Range} and associated animations is that total torques and torque components vary greatly over short timescales.  Total torques are usually positive but occasionally reverse sign in a limited radial range as a result of large variability in low-order $m$ contributions to the total torque. 

Significant bursts in low-order (particularly $m=$ 2, 3) gravitational torque components sometimes range from ~11 to 30 au or more (see Figure 16 and the animation linked to that figure). 
Sudden increases in $m = 1$ and, to some extent, $m = 2$ and $m = 3$ torques sometimes range over a significant fraction of the disk's radial extent.   Torque outbursts arising from higher-order symmetries, like $m$ = 4 and 5,  are more local, but each still shows radial extents of 6--30 au with the radial range of efficacy and extent changing with time, even on short timescales.  An example of this is seen in the the upper rightmost set of panels in Figure~\ref{fig:BigPicture} (17.85 ORPs), where the blue curve for $m = 4$ shows an amplitude burst extending over more than 20 au. 

Although instantaneous torques vary considerably from their time-averaged values, the overall characteristics of broad humps in the surface density near 9, 11, and 14 are preserved. 
 
\subsection{Effective $\alpha$}

It has long been recognized that molecular viscosity alone is not sufficient to account for angular transport in accretion disks \citep[e.g.,][]{Durisen_2011}. Some additional process must augment or dominate molecular viscosity.  \cite{shakura1973} proposed turbulence in the gas as the source of this viscosity.  Assuming subsonic turbulence and an upper limit on the size of eddies, they proposed  the ansatz
\begin{equation}
\nu = \alpha c_s H,
\label{eq:viscous_coefficient}
\end{equation}
where 
$\nu$ is the coefficient of viscosity responsible for carrying angular momentum outward, $c_s$ is the sound speed, and $H$ is the disk scale height.  The free parameter $\alpha$ provides the link between known quantities and turbulent viscosity.  Given this equation, $\alpha$ and thus $\nu$ are locally defined. 

Following the development of  \cite{gammie2001}
\citep[see also][]{lodato2004, boley2006, Michael2012}, an effective $\alpha_{\rm{\rm eff}}$ arising from gravitational stresses alone can be written as
\begin{equation}
\alpha_{\rm{\rm eff}}(r) = \left| \frac{{\rm d} \ln \Omega}{{\rm d} \ln r} \right|^{-1} 
\frac { \langle T_{r\phi}^{grav} \rangle} { \int_z \rho c_s^2 {\rm d}z },
\label{eq:alpha2}
\end{equation}
where $\Omega$ is the azimuthally averaged rotation speed, $T_{r\phi}^{grav}$  is the  gravitational stress tensor, and $\rho$ is the volume density; angle brackets indicate azimuthal averages of vertically integrated stresses.  The integrated divisor is also azimuthally averaged. 

The gravitational stresses of Equation \eqref{eq:alpha2} can be evaluated in a straightforward manner using the gravitational torque of Equation \eqref{eq:cgravzeq}, i.e., 
\begin{equation}
 T_{r\phi}^{grav}\left( r \right)   \; = \; \frac{1}{2\pi r^2}C_Z \;
                                 = \; - \frac{1}{2\pi r^2} \int \rho\frac{\partial\Phi}{\partial\phi} {\rm d}V.
\label{eq:cgravzeq2}
\end{equation}

Using a 2D shearing box approach, which formally included both gravitational and hydrodynamic stresses and enforced balance between heating and cooling, Gammie formulated an effective $\alpha$ \citep[see also][]{pringle1981} expressed in 3D by Eq. (21) of
\citet{Bethune_Latter_Kley2021} as
\begin{equation}
\alpha_{\rm eff}(r) =   \Big| \frac{ {\rm d }\log \Omega} {{\rm d} \log r}  \Big|^{-1}   [ \gamma(\gamma - 1) \beta]^{-1},
\label{eq:Bethune_Pred}
\end{equation}
\noindent
where $\beta = \Omega t_{\rm cool}$ is the normalized local cooling time and $\gamma$ is the adiabatic index. 
In this formulation, for fixed $\gamma$ in a Keplerian disk, $\alpha_{\rm eff}$ is a function only of $\beta$ and thus the effective viscosity depends only on the cooling time.   As described in Section \ref{subsection:hydrodynamics}, for the temperature range in our simulations the gas is well approximated by an adiabatic index $\gamma = 5/3$.


\begin{figure}[htb!]
\vspace{-10pt}
\centerline{\includegraphics[width=1.0\textwidth]{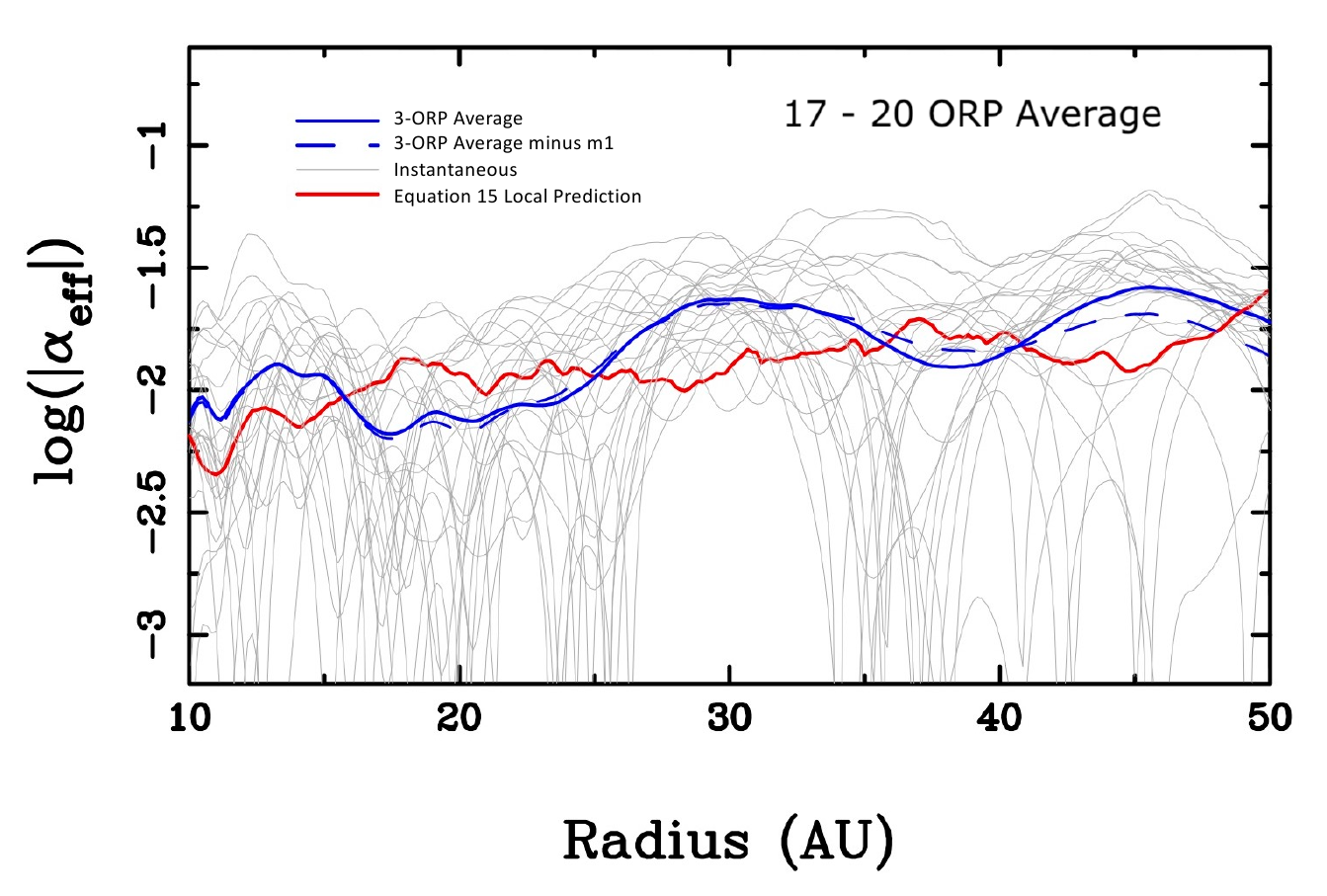}}
\caption{Effective $\alpha$  for the converged disk.  Light gray lines show instantaneous $\alpha_{\rm eff}$ at a subset of 25 equally spaced time steps  
between 17 and 20 ORPs calculated using Equations~\eqref{eq:alpha2} and \eqref{eq:cgravzeq2}.
The time average of instantaneous $\alpha_{\rm eff}$ is displayed by the solid blue line, while the dashed blue line displays the time average minus contributions from the $m=1$ Fourier component.  Predictions for $\alpha_{eff}$ from Equation~\eqref{eq:Bethune_Pred}, are shown by the thick red line. }
\label{fig:alphaplot}
\end{figure}

Figure \eqref{fig:alphaplot} shows time-averaged $\alpha_{\rm eff}(r)$ for the converged disk, determined using the gravitational stresses of Equations \eqref{eq:alpha2} and \eqref{eq:cgravzeq2}, averaged at 240 equally spaced times between 17 and 20 ORPs along with a subset of 25 instantaneous measures of $\alpha_{\rm eff}(r)$ equally spaced in time. Because the $m=1$ Fourier component of the mass distribution does not directly arise from GIs (\S\ref{subsection:Transport}), we show time-averaged $\alpha_{\rm eff}(r)$ inclusive of all Fourier components and minus the $m=1$ Fourier component. For comparison, local predictions of Equation~\eqref{eq:Bethune_Pred} averaged over the same limiting times and using instantaneous $\beta$ from the ACDC are shown.   

Based on evolutionary lifetimes of protostellar disks, Hartmann et al. (1998) estimated $\alpha \approx  0.01$ at disk radii between 10 and 100 au.  Depending on disk parameters, saturated GIs provide transport at rates such that $10^{-2} < \alpha_{\rm eff} < 1$ \citep{2016-ARAA-Kratter-Lodato}.   Our time-averaged curve (solid blue line) falls between $\sim 10 ^ {-2.2}$ and $10 ^ {-1.6}$, consistent with this result.  In contrast, {\em instantaneous} $\alpha_{\rm eff}$ vary by greater than an order of magnitude over much of the disk and are, at times, considerably lower than the time-averaged values.  This reflects the great deal of variability in the torques discussed in the previous section. The deep minima of instantaneous $\alpha_{\rm eff}$ seen in Figure \eqref{fig:alphaplot} occur because the $\alpha_{\rm eff}$  are subject to sign reversals due to sign reversals in the torque. When the torques pass through zero, one sees deep minima in $\log(|\alpha_{\rm eff}|)$.  Hence, reader needs to realize that, although the averaged $\alpha(r)$ values look relatively smooth in Figure~\ref{fig:alphaplot}, there are several sign reversals leading to several radial ranges of {\it negative} $\alpha$. This does not much resemble the picture of a simple $\alpha$-disk.

ACDC torque-based $\alpha_{\rm eff}$ and the local dissipation predictions of Equation~\eqref{eq:Bethune_Pred} track each other over the full radial range of the disk and are intertwined such that they precisely agree at six different radii and differ by up to a factor of two between these radii of agreement. Better agreement is seen in the outer disk when $m = 1$ torques are not included. 

Between 16 and 25 au, the local dissipation curve is above the torque-based curve, suggesting that a stress or energy transport term is missing.  Where does this energy go? It seems to heat the disk inside 16 au and the region outside 25 au.  How important is it? It is about 50\% of the heating by gravitational stress. 

We suspect that this energy flow is due to Reynolds stresses caused by the correlated deviations from mean circular motion in global spiral modes. In other words, the low-order global spiral waves, especially $m =$ 2--4, are transporting energy nonlocally. It is also possible that there is radiative radial transport caused by the large temperature gradients associated with shocks. These sources of energy flow are difficult to tease out of our current simulations. The spiral modes erupt in a chaotic manner over the whole low-$Q$ part of the disk. The deviations of the red and blue curves are the combined effect over time of many spiral modes appearing and disappearing. This is truly gravitoturbulence, but dominated by fluctuating large-scale modes. It would be useful to study the Reynolds stresses and radial radiative transport due to erupting global spirals in more detail. Such an analysis goes beyond the scope of the current paper, but Figure~\ref{fig:alphaplot} suggests that such a study might be interesting and  fruitful.

We further note that 16, 25, and 42 au, radii where both local and torque-predicted $\alpha_{\rm eff}$ curves agree in Figure~\ref{fig:alphaplot}, are the same radii where torque-predicted and measured $dM/dt = 0$ in Figure~\ref{fig:torqueplot}.

Total torques at 30 au, where time-averaged torque-based $\alpha_{\rm eff}$ peak,  never drop to zero and always are fairly large.  This can be seen in the instantaneous curves of Figure \eqref{fig:alphaplot}.   
As seen in Figure 19, there are frequent $m=$ 2, 3, and 4 outbursts centered around 30 au.  In addition, as shown in Figure~\ref{fig:Mass_on_Cylinders}, a persistent ``bump'' is seen in the radial mass distribution at this same radius.

\subsection{Ring-like Structures}\label{subsection:rings}


\begin{figure}[htb!]
\vspace{-10pt}
\centerline{\includegraphics[width=0.6\textwidth]{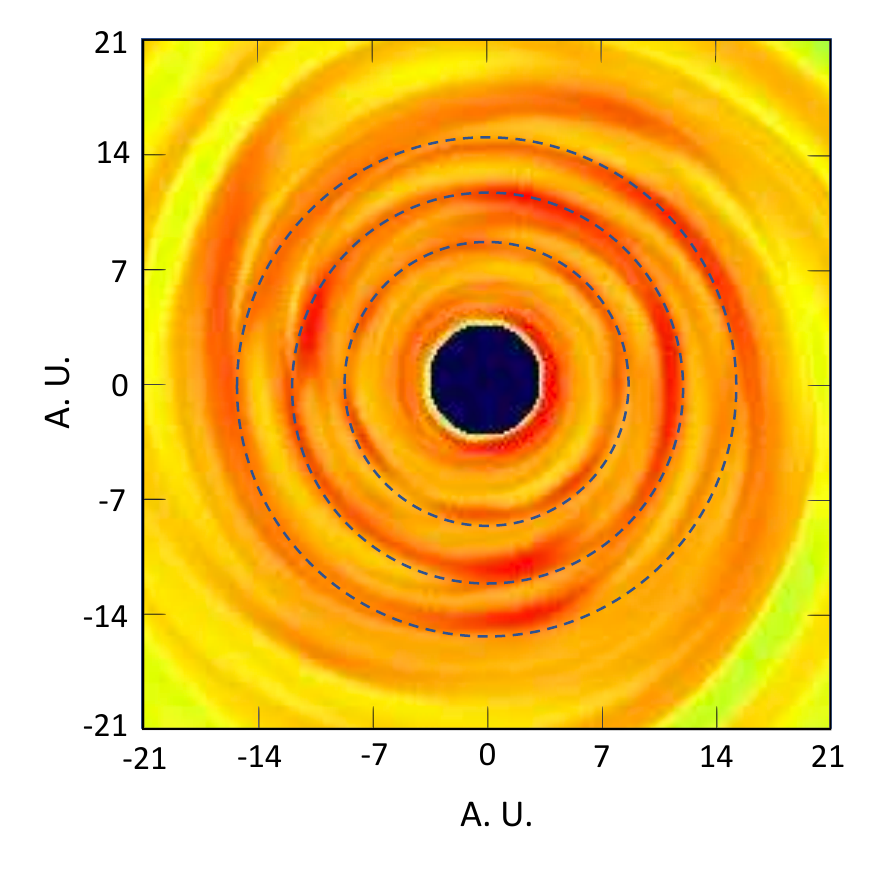}}
\caption{
The central regions of Figure \ref{fig:FullDisk} 
with locations of the 8, 11, and 14 au ``ring'' structures shown.  Careful inspection reveals 
tightly wound spiral arms originating in the 11 au feature.  This radius corresponds with the ILR  of $m = 2$, 4, and 6 modes. }
\label{fig:CentralDisk}
\end{figure}


\begin{figure}[htb!]
\vspace{-10pt}
\centerline{\includegraphics[width=0.5\textwidth]{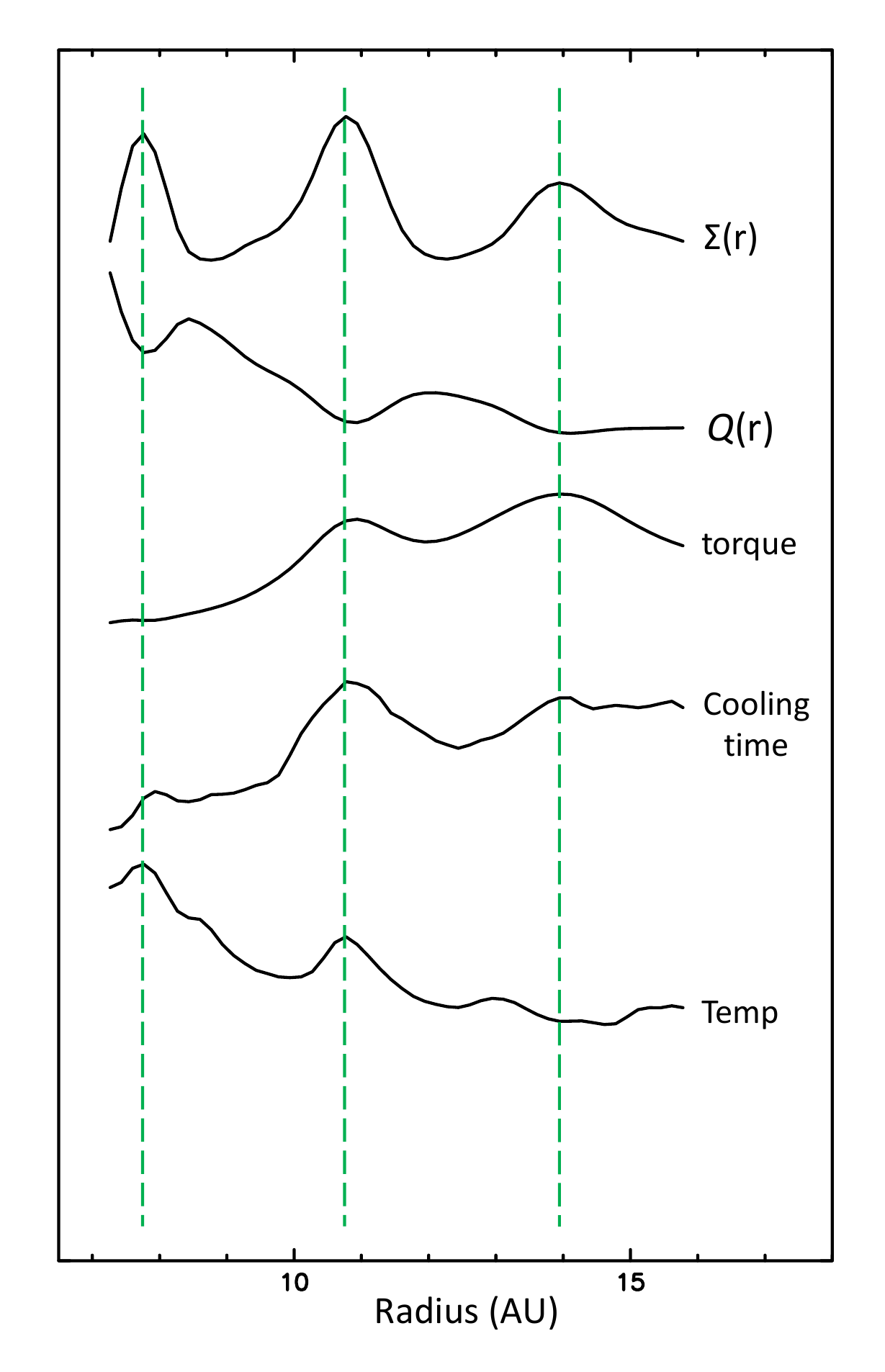}}
\caption{Comparison of ring centers (dashed line) with local extrema of the surface density 
(Figure~\ref{fig:sigmaplot}), $Q$ (Figure~\ref{fig:qplot}), time-averaged gravitational torques (Figure~\ref{fig:torqueplot}), cooling times (Figure~\ref{fig:Cool_Times}), and temperatures (Figure~\ref{fig:Temperatures}).  Curves are rescaled and vertically offset to aid readability.
}
\label{fig:rings}
\end{figure}

Prominent and persistent ring-like structures are present at 8, 11, and 14 au in the converged disk (locations shown by dashed lines in the Figure~\ref{fig:CentralDisk}).  
These rings contain ``excess” masses  of $\sim 6M_{\rm J}$, $18M_{\rm J}$, and $10M_{\rm J}$, respectively, and correlate strongly with physical characteristics of the disk as shown in Figure \ref{fig:rings}.  In particular, ring radii correspond with local minima in $Q$, and local maxima in the surface density, time-averaged gravitational torques, cooling times, and  temperatures.  Semipersistent dense clumps are present in each of the rings.  
Time animations of the radially averaged surface density $\Sigma(r)$ and the 31-32 au bump in the radial mass distribution show cyclical variations with time (see animation link in Section 3.5).


\begin{figure}[htb!]
\centerline{\includegraphics[width=0.8\textwidth]{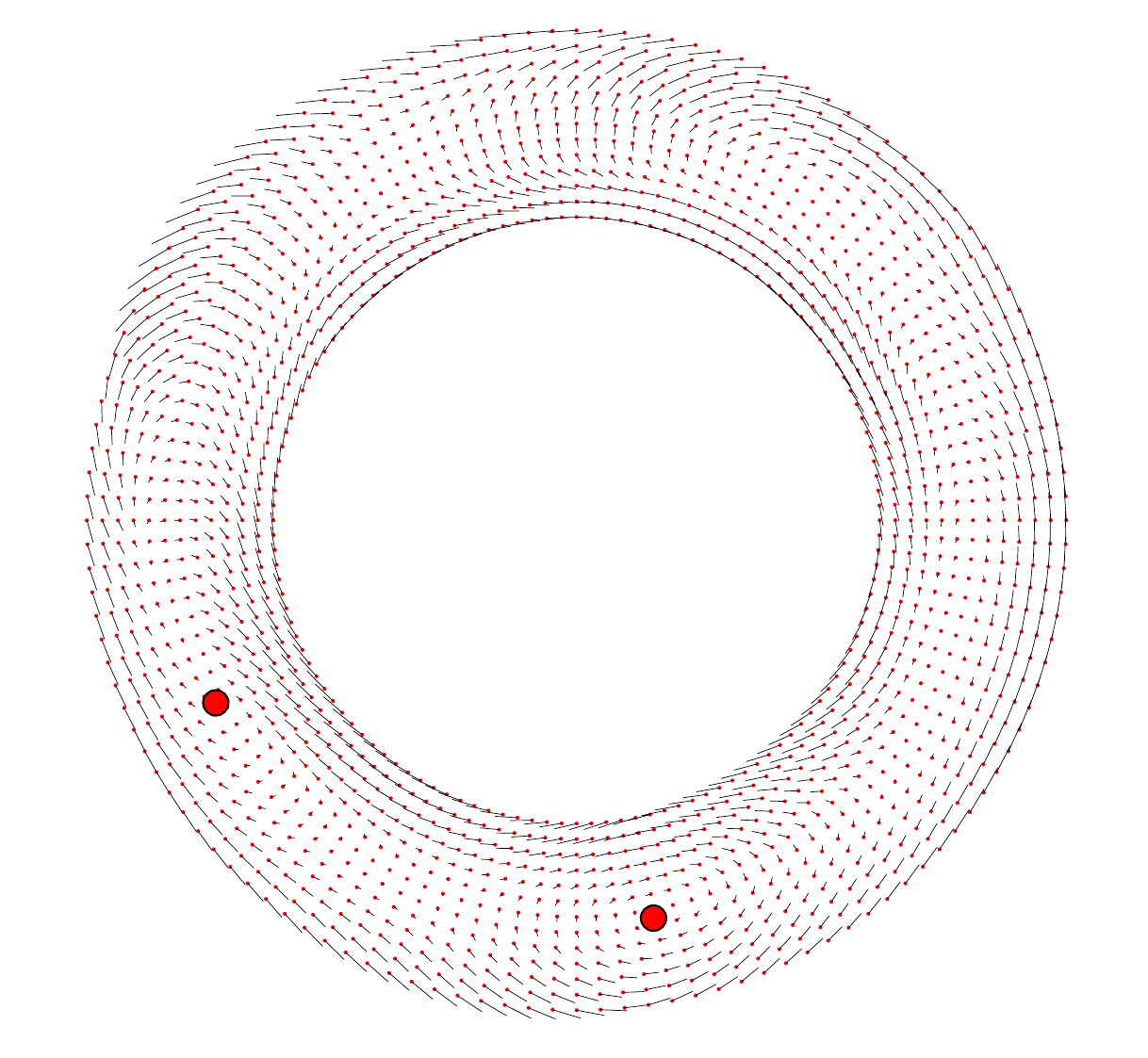}}
\caption{
Velocity field at $t=$ 20 ORPs between 9.8 and 11.8 au relative to the azimuthally averaged velocity at 10.8 au, the center of the 11 au ring. The radial appearance of the annular region has been stretched to facilitate visibility.  Velocity vectors are shown with small red circles at their heads.  The sense of bulk rotation for the ring is counterclockwise.    Strong noncircular motions and vorticities, marked by large red circles, are prominent.  }
\label{fig:VelocityField}
\end{figure}

Ring locations correspond with several low-order orbital resonances.  The 8 au ring serves as the ILR of an $m = 2$ mode and also the inner $Q$-barrier.  
Torques arising from $m > 1$ structures are negligible interior to the 8 au ring (Figure~\ref{fig:Torque_Range}), but this is not the case for $m = 1$.  In the parlance of \citet{Durisen_etal_2005Icarus}, the 8 au ring corresponds with the “active boundary ring” (ABR), because it occurs at the boundary between GI-active and GI-inactive regions and because it displays active nonaxisymmetric dynamics.
 The 11 au ring serves as the ILR of $m = 2$, 4, and 6 modes, while the 14 au ring serves as the ILR of $m = 3$, 4, and 6 modes, as well as the CR radius of $m = 4$ and $m = 6$ modes.

Similar ring features have been seen in previous 3D hydro simulations of self-gravitating PPDs carried out by our group and collaborators
\citep{pickett1996, pickett2003,  mejiaphd2004,  mejia2005, Durisen_etal_2005Icarus, caiphd2006,  boley2006, boley2007_bdnl, cai2006, cai2008, Michael2012, Steiman2013, desai2019}.  These studies find that rings form early in the simulation, well before disks settle into their asymptotic state, and persist throughout the settled asymptotic state. 
More recently, a 2D study of an eccentric spiral instability in a self-gravitating disk with cooling by \citet{Ring_Formation_LiEtal_2021ApJ}  found that a trapped one-arm instability forms early in the simulation and evolves into a set of axisymmetric rings.  Their disk was not subject to GIs, but, on a deeper level, the dynamics involved may be similar.     

The rings found in this study correspond with local time-averaged torque maxima.  Because both $\Sigma$ and temperatures are local maxima at ring locations, rings are also pressure maxima \citep[see also ][]{Carrera_etal_2021AJ}.  
Assuming that these pressure maxima rings are, roughly speaking, in radial force balance, the rotation curve $\Omega(r)$ will be supra-Keplerian on the inner slope and sub-Keplerian on the outer slope of rings.  We have confirmed that this is the case in the simulation. 

Measured fluid flows in the 8 - 14 au region that encapsulates the rings are complicated and nonsteady, with distinct radial flows and anticyclonic vortices around local clumps, as shown in Figure~\ref{fig:VelocityField}.  Velocity vectors are shown with small red circles at their heads, and the general bulk rotation is counterclockwise.  Large red circles represent the centers of two vortices. These vortices are almost certainly caused by the Rossby wave instability \citep{Li_etal2000ApJ,Li_etal2001ApJ}.  Disk vortices have long been heralded as promising routes for planet formation owing to their ability to trap significant amounts of small particles  \citep{Heemskerk_etal1992, Adams_Watkins_1995ApJ, Klahr_Henning_1997Icarus, Lovelace_etal1999ApJ, Li_etal2000ApJ, Li_etal2001ApJ, Klahr_Bodenheimer_2003ApJ, Klahr_2004ApJ, Klahr_2008PhST, Lyra_etal2009AA, Lovelace_Hohlfeld_2013MNRAS, Meheut_Lovelace_2013MNRAS, Lovelace_Romanova_2014FlDyR,  Klahr_Hubbard_2014ApJ, Lovelace_Li_2014APS, Li_Li_2015ASPC, Lyra_Umurhan_2019PASP, Luane_etal2020ApJ, Klahr_Schreiber_2020ApJ, Pfeil_Klahr_2021ApJ}.  

\subsection{Grid Convergence and $l_{\rm max}$}

Convergence in this study is quite good.  The results cited above permit an assessment of appropriate azimuthal grids for the problem under consideration and similar problems.  

The $l_{max}=512$ disk (``converged disk'') simulation has clearly achieved grid convergence in the $\phi$-direction and settled into an asymptotic phase before 17 ORPs for regions interior to $\sim$50--52 au.  Most of the analyses presented in this paper have been performed in the 17--20 ORP time window.   All four simulations settled into similar surface densities $\Sigma(r)$ (Figure~\ref{fig:sigmaplot}), with differences of 25\% or less between 10 and 52 au in their thermodynamic states, as represented by their asymptotic $Q(r)$ (Figure~\ref{fig:qplot}), where $l_{max}=64$ differs from the 512 disk by 15-30\% and the 128 disk by roughly half that amount. In contrast, $Q$ for the 
256 disk varies from the 512 disk by an amount comparable to, or less than, the size of cyclical variations in $Q(r)$

Gravitational torques for the 256 and 512 disks are dominated by $m =$ 1 - 6 Fourier components.  Power spectra of these components (Figure~\ref{fig:amplot}) and the resultant time-averaged torques display very good convergence for this range of Fourier components, with minimal differences between the 256 and 512 simulations.  Thus, for a quick examination of radially averaged $Q(r)$, surface densities, 
and torques, a modestly low resolution of $l_{\rm max}=128$ may suffice in disks dominated by only the lowest-order spiral structure. For most purposes, a simulation with $l_{\rm max} = 512$ appears to offer no material advantage over an $l_{\rm max} = 256$ calculation.

It has been argued that prompt fragmentation is more likely to occur as the resolution of 3D simulations increases. For the range of resolutions presented here, there is no evidence that this is the case. It must be noted, however, that the full time duration of the simulations reported here is only $\sim$ 5000 yr, with the converged asymptotic state followed for $\sim$ 800 yr.  Therefore, depending upon one's definition of ``prompt", these simulations cannot preclude or confirm the possibility that limited resolution (however that is defined) might prevent fragmentation in some instances.

Paper II and this work both examined the same disk using identical initial conditions and the same hydrodynamics code, albeit with two important modifications.  Specifically, this paper includes the implementation of a subcyling approach to better control heating and cooling (Section 2.2) and the inclusion of an indirect potential approach to self-consistently account for star – disk interactions  (Section 2.3).  In contrast to Paper II, where convergence was lacking, great convergence is found here.  
The state to which a nonfragmenting disk settles is fairly sensitive to the accurate treatment of radiative losses, leading to higher (and more nearly constant) $\alpha$ values, constant $Q$ over the optically thick disk (much in line with approximations used by others), higher $Q$' values in the optically thin regions (at least for our disk).   The results of this work demonstrate the importance of doing the radiative physics well. In addition, the greater prominence of $m = 1$ here, compared with our earlier papers, indicates that one needs to include star--disk interaction \citep[see also Section 6.5 of][]{Elbakyan_2023MNRAS}.
\hspace{15em}
 
\section{Discussion}

The existence of pressure maxima at ring centers may have important implications for 
planetesimal/planet formation.  
\citet{Weidenschilling1977MNRAS} was the first to point out that solid particles orbiting in a gaseous disk drift radially in the direction of a radial pressure gradient 
\citep[see also][]{Cuzzi_Dobro_Champney1993Icarus}.  
\citet[][hereafter HBa, HBb]{hag2003a,hag2003b} subsequently studied the motions of small solids in the vicinity of local pressure enhancements of a gaseous nebula.  They
showed that the combined effects of gas drag and pressure gradients lead solids to accumulate at the locations where the pressure of the gas maximizes. 

We can envision a path where solid particles accumulate in the rings owing to radial drift toward pressure maxima in rings.  These particles can then be trapped by vortices within the rings which could, in turn, accelerate the growth of protoplanets.  \citet{Durisen_etal_2005Icarus} used the results of HBa and HBb to estimate drift times in rings that appeared in an earlier disk simulation, leading them to suggest that even if instabilities due to disk self-gravity do not produce gaseous protoplanets directly, they may create persistent dense rings that are conducive to accelerated growth of gas giants through core accretion. 

The coincidence of strong resonances with rings, as described in Section 3.4, is an argument for the probable role of resonances with GI waves in ring formation \citep{Durisen_etal_2005Icarus}.   
Eccentric modes, corresponding to perturbations with azimuthal wavenumber $m = 1$, have also received particular interest in the context of PPDs because of their global nature. 
A large corpus of work has examined the development and sustenance of these modes in fluid disks. 
These have shown that almost any disk with a realistic density profile can sustain long-lived eccentric modes \citep{Lee_etal_2019b} and that, once initiated, a global eccentric mode can grow its amplitude via the sling mechanism that amplifies an eccentric perturbation through the wobble of the central star and instantaneous cooling
\citep{AdamsRuden1989, Shu_etal_1990ApJ,Lin_2015MNRAS.448.3806L}.  Ring formation then follows via angular momentum exchange with an unstable eccentric mode 
\citep{Lubow_1991ApJ,Ogilvie_2001MNRAS,Lee_etal_2019a,Lee_etal_2019b, Ring_Formation_LiEtal_2021ApJ}.  

For all these reasons, we expect ring formation to be a common product of PPD evolution and it may play an important role in giant planet formation. 

\subsection{Rings and Closely Spaced Giant Planets}

Given the discussions above and in Section \ref{subsection:rings}, we speculate on a possible mechanism for forming closely spaced giant planets.  The Nice model for the early dynamical evolution of the solar system \citep{2005NICE_Model1,2005NICE_Model2,2005NICE_Model3,Levison_etal2008Icarus,Nesvorny_2012AJ,Brasser_2013Icarus,Nesvorny_2013} proposed the migration of the giant planets from an initial compact configuration into their present positions, long after the dissipation of the solar PPD.  Among other things, it successfully explains the late heavy bombardment of the inner solar system, the formation of the Oort Cloud, and the existence of populations of small solar system bodies.  A critical aspect of the Nice model is that the four giant planets were originally in much more closely spaced orbits than today. 

 At the earliest phases, PPDs have spiral waves. If the growth of ring-like enhancements in $\Sigma(r)$ is a natural process and persists long enough, then the disk will have true rings left after GIs largely shut down.  Extending this line of thought, if the azimuthal mass concentrations (clumps) along with associated sustained vortices in these rings persist, this could lead to the growth of multiple Jovian and/or ice giant planets within a relatively small radial extent.  Interactions between these newly formed massive planets would then redistribute them into a more stable orientation and move icy planetesimals into the inner disk. 

We note that  issues may exist with this scenario, in particular, the delayed action between the ring and planet formation in closely spaced rings. The spacing of our rings may give highly unstable planets. Of course, with different disk parameters and following the whole rapid infall phase, any possible set of rings might look quite different. In addition, the structure of the inner disk could differ a lot if other methods for generation of turbulence were included in regions too hot to produce GIs. These are serious concerns and should be mentioned as caveats to this idea.

The \citet{Durisen_etal_2005Icarus} paper about a hybrid theory of planet formation attributes the rings in part to ILRs of high-$m$ modes, and the authors proposed that heating of the inner disk might be due to dissipation in the innermost disk of GI waves generated in the central low-$Q$ region of the disk. Alternately, the rings may be due instead to Rossby wave Instabilities, which in the nonlinear regime produce vortices, as shown by \citet{Ring_Formation_LiEtal_2021ApJ}.  In reality, perhaps both mechanisms operate. It is also interesting that broad surface density bumps appear and persists near 30 and 48 au in our simulation. Hence, a GI-active disk can generate long-lived radial structures in which solids may accumulate, not just near edges.

\section{Summary and Conclusions}

We present results of a 3D grid-based radiative hydrodynamics convergence study of a 0.07 $M_\odot$ protoplanetary disk subject to GIs surrounding a 0.5 $M_\odot$ star.  The disk is evolved with a significantly improved radiative transport scheme using realistic dust opacities.   This work supersedes the work of \citet{Steiman2013}.   Both works examined the same disk, but here with important improvements to the hydrodynamics code, including the implementation of a subcyling approach to better control heating and cooling and the inclusion of an indirect potential approach to self-consistently account for star -- disk interactions that inevitably displace the star from the center of mass. 

Our goals include determining cooling times experienced by the disk, characterizing the spiral density perturbations produced by the GIs and the gravitational torques produced by these structures, understanding disk processes on both time-averaged and instantaneous timescales (time averaging can hide dynamically important time variabilities that affect both the short- and long-term properties of the disk), evaluating the level to which transport can be represented as a local or nonlocal process, and understanding ring-like structures in the inner disk and their possible role in planet formation.

Four simulations were conducted and followed through to the time when the disks have settled into an asymptotic state where heating and cooling are roughly in balance.
These simulations were identical except for the number of azimuthal computational grid points, thus allowing for calculations to establish mesh convergence in the $\phi$-direction.

\begin{enumerate}
\item
The primary messages of this work are that GI-active disks are awash in vigorous dynamic behavior, sloshing around considerably more mass on short timescales than the net transport, exhibiting rapidly changing gravitational torques with recurrent amplitude bursts.  Edges of various sorts, e.g., optically thick to optically thin, variable cooling times, real physical edges, radially varying $Q$ at those edges, rings, etc., are important.  Edges affect many physical processes that vary in time and space and thus underlie the global nature of GI-active PPDs.
\item
Accurate treatment of radiative cooling is critically important. With the heating and cooling limiters used in Paper II, the disk remained significantly too hot, leading to higher $Q$ values, higher optical depths, and mostly incorrect cooling times. Moreover, there was no convergence in Paper II even with $l_{\rm max} =$ 512.  Results in this paper, based on subcycling of the energy equation for radiative cooling, are thermally smoother and converge quickly as the azimuthal resolution is increased.
\item
All simulations settled into similar radial surface density $\Sigma(r)$ and Toomre $Q(r)$ profiles.  $\Sigma(r)$ is well fit by an exponential profile over most of the disk.  Superposed on this general trend are persistent local ring-like maxima at $\sim8$, 11, and 14 au that  contain ``excess” masses of   $\sim6M_{\rm J}$, $18M_{\rm J}$, and $10M_{\rm J}$, respectively.  In addition, broad fluctuating but persistent bumps in the radial mass distribution are seen around 30 and 48 au.  The former location corresponds with a maxima of the time-averaged gravitational torques.  These torques serve to maintain the density maximum.  The latter is associated with a strong one-arm spiral feature in the outer disk.
\item
The ring-like features correspond with local torque and pressure maxima.  Fluid flows in the region encapsulating the rings are complicated, with distinct radial flows and anticyclonic vortices around local clumps within in the rings.  
\item
Two distinct radial regions were found with essentially constant $Q$, with each displaying different convergence characteristics.  {\it Region 1} lies between $\sim$ 11 and 32 au and is defined by $Q \approx 1.4$.  {\it Region 2} is bounded by 40--50 au and is characterized by $Q \approx 2.1$.  Between these regions, $Q$ increases in a roughly linear fashion with radius. These two regions were initially identified and defined by the $Q$ behavior only. Further analysis determined that Region 1 was optically thick, Region 2 was optically thin, and the transition between them contained the vertical $\tau =1$ transition region.
\item
{\it Region 1} spans the fully optically thick portion of the disk.  Cooling times peak at the inner edge of the region with $t_{\rm cool}/P_{\rm orb} \sim 33$ and decrease to $\sim 14$ orbital periods at the outer limit of the region.   Optically thick spiral waves, arising from GIs, are embedded in the optically thick background of Region 1. Shock-heated cells are common in locations that strongly correlated with the spiral waves.  Torques in Region 1 are dominated by the $m$ = 2 -- 5 Fourier components of the azimuthal mass distribution.  
\item
{\it Region 2} spans the fully optically thin portion of the disk.  Here only low-order Fourier components appear dynamically important. Optically thick spiral waves dominated by $m = 1$ with moderately strong $m = 2$ are embedded in the optically thin background of Region 2.  Local cooling times are smaller than in Region 1, dropping from $t_{\rm cool}/P_{\rm rot}$ $\sim 10$ at the inner edge to 5 at the outer edge.
\item
We find several low-$m$ (2--6) modes in Region 1 and the transition region, with ILR ranging from 8 to 32 au and OLR ranging from 18 to 40 au.   Gravitational torques arising from these modes vary in strength by up to a factor of 10 in a roughly cyclical manner.  We attribute this to recurrent swing-amplified bursts that cycle on approximately the local dynamical time. This activity is particularly strong over the radial range between 8 and 18 au, a region dominated by ring structures between 8 and 14 au.  In addition to these persistent well-defined modes, numerous transitory {\it ephemeral modes} are found.  While these have well-defined orbital resonances (ILR,CR,OLR), they are not persistent throughout the simulation, but rather come and go.  
\item
The agreement between $\alpha$ values due to gravitational torques alone and expectations from measured cooling times is fairly good and gives values $\sim10^{-2}$ and hence mass transport rates 
$\sim \pm 10^{-7} M_\odot$ yr$^{-1}$ 
and evolutionary time scales $\sim10^6$ yr.   Deviations of a factor of two about this agreement suggest that some stresses or transport mechanism are not accounted for. What this and our earlier papers have shown, however, is that the disk behavior is not at all well characterized by a simple $\alpha$-disk. This GI-active disk produce bands of inflow and outflow of mass, associated with significant persistent and/or episodically eruptive spiral structures. The disk is highly dynamic on a range of time scales in a nonlocal manner.
\end{enumerate}

\bibliography{general}
\bibliographystyle{aasjournal}

\end{document}